\documentclass[preprint]{aastex63}
\usepackage[utf8]{inputenc}
\usepackage{natbib}
\usepackage{color}
\usepackage{textcomp}
\usepackage{graphicx}
\usepackage{rotating}
\usepackage{hyperref}
\usepackage[all]{hypcap}
\usepackage{mathrsfs}
\usepackage{amsmath}
\usepackage{amssymb}
\usepackage{float}
\usepackage{gensymb}
\usepackage{subfigure}
\usepackage{lineno}
\usepackage[version=4]{mhchem}
\hypersetup{
    bookmarks=true,         
    unicode=false,          
    pdftoolbar=true,        
    pdfmenubar=true,        
    pdffitwindow=false,     
    pdfstartview={FitH},    
    pdftitle={My title},    
    pdfauthor={Author},     
    pdfsubject={Subject},   
    pdfcreator={Creator},   
    pdfproducer={Producer}, 
    pdfkeywords={keyword1} {key2} {key3}, 
    pdfnewwindow=true,      
    colorlinks=true,        
    linkcolor=blue,         
    citecolor=blue,         
    filecolor=magenta,      
    urlcolor=cyan           
}
\usepackage{xcolor}
\usepackage[detect-all,table-format=2.3]{siunitx}
\newcommand{\p}[1]{{\color{red}{#1}}}

\begin{document}
\title{Data-constrained magnetohydrodynamic simulation of global solar corona including solar wind effects within $2.5~R_\odot$}

\author[0009-0003-0312-2513]{Yihua Li}
\affil{School of Astronomy and Space Science and Key Laboratory of Modern Astronomy and Astrophysics, Nanjing University, Nanjing 210023, China}

\author{Guoyin Chen}
\affil{School of Astronomy and Space Science and Key Laboratory of Modern Astronomy and Astrophysics, Nanjing University, Nanjing 210023, China}
\affil{Rosseland Centre for Solar Physics, University of Oslo, Oslo, Norway}

\author{Jinhan Guo}
\affil{School of Astronomy and Space Science and Key Laboratory of Modern Astronomy and Astrophysics, Nanjing University, Nanjing 210023, China} 

\author[0000-0002-9293-8439]{Yang Guo}
\affil{School of Astronomy and Space Science and Key Laboratory of Modern Astronomy and Astrophysics, Nanjing University, Nanjing 210023, China} \email{guoyang@nju.edu.cn}

\author{Hao Wu}
\affil{School of Astronomy and Space Science and Key Laboratory of Modern Astronomy and Astrophysics, Nanjing University, Nanjing 210023, China}
\affil{Centre for mathematical Plasma Astrophysics, Department of Mathematics, KU Leuven, Celestijnenlaan 200B, B-3001 Leuven, Belgium}

\author{Yuhao Huang}
\affil{School of Astronomy and Space Science and Key Laboratory of Modern Astronomy and Astrophysics, Nanjing University, Nanjing 210023, China} 

\author{Xin Cheng}
\affil{School of Astronomy and Space Science and Key Laboratory of Modern Astronomy and Astrophysics, Nanjing University, Nanjing 210023, China}

\author[0000-0002-4978-4972]{M. D. Ding}
\affil{School of Astronomy and Space Science and Key Laboratory of Modern Astronomy and Astrophysics, Nanjing University, Nanjing 210023, China} 

\author[0000-0003-3544-2733]{Rony Keppens}
\affil{Centre for mathematical Plasma Astrophysics, Department of Mathematics, KU Leuven, Celestijnenlaan 200B, B-3001 Leuven, Belgium}

\begin{abstract}
Total solar eclipses (TSEs) provide a unique opportunity to observe the large-scale solar corona. The solar wind plays an important role in forming the large-scale coronal structure and magnetohydrodynamic (MHD) simulations are used to reproduce it for further studying coronal mass ejections (CMEs). We conduct a data-constrained MHD simulation of the global solar corona including solar wind effects of the 2024 April 8 TSE with observed magnetograms using the Message Passing Interface Adaptive Mesh Refinement Versatile Advection Code (MPI-AMRVAC) within $2.5~R_\odot$. This TSE happened within the solar maximum, hence the global corona was highly structured. Our MHD simulation includes the energy equation with a reduced polytropic index $\gamma=1.05$. We compare the global magnetic field for multiple magnetograms and use synchronic frames from the Solar Dynamics Observatory/Helioseismic and Magnetic Imager to initialize the magnetic field configuration from a magneto-frictionally equilibrium solution, called the Outflow field. We detail the initial and boundary conditions employed to time-advance the full set of ideal MHD equations such that the global corona is relaxed to a steady state. The magnetic field, the velocity field, and distributions of the density and thermal pressure are successfully reproduced. We demonstrate direct comparisons with TSE images in white-light and Fe XIV emission augmented with quasi-separatrix layers, the integrated current density, and the synthetic white-light radiation, and find a good agreement between simulations and observations. This provides a fundamental background for future simulations to study the triggering and acceleration mechanisms of CMEs under solar wind effects.
\end{abstract}

\section{Introduction} 
\label{sec:intro}
Coronal mass ejections (CMEs) are among the most important events observed within the solar system. These intense eruptions release large amounts of magnetized plasma from the Sun into the interplanetary space, significantly influencing the heliospheric environment and magnetosphere of the Earth \citep{chen2011review,guo2017review,cheng2017review}. Coronagraphs and total solar eclipses (TSEs) provide observations of CMEs and large-scale coronal structures. In particular, a TSE raises a rare opportunity to directly observe fine structures within the large-scale corona using ground-based instruments \citep{boe2020TSE,habbal2011arXivTSE,habbal2021ApJTSE}.

Magnetohydrodynamic (MHD) simulations aiming to investigate the eruption and propagation of magnetic flux ropes (MFRs) often neglect the influence of the solar wind at relatively high altitudes, particularly when the selected simulation domain is large in spherical coordinates \citep{guo2021rbslsphere,li2025rbsl}. In instances where the simulated domain extends over several solar radii (i.e., $2-6~ R_\odot$), it is significant to account for the effect of the steady solar wind flow on the distribution of coronal matter and eruption velocity, which subsequently impacts the topology and configurations of the magnetic field \citep{fan2017mhd,fan2022improved,guo2023MFRbirth-death}. In MHD simulations, the initial potential field configuration, usually potential field source surface (PFSS) model \citep{Schatten1969pfss,Altschuler&Newkirk1969pfss,toth2011pfss,nikolic2019pfsssolutions,porth2014mpi,xia2018mpi}, does not consider the solar wind effects adequately, potentially influencing the velocity and energy evolution of the CME \citep{fan2016modeling}. Moreover, commonly utilized global magnetic field models fail to adequately incorporate the role of the solar wind in the corona and raise a general open-flux problem \citep{linker2017open, rice2021global, asvestari2024openfield}, which reveals that the open magnetic flux predicted by PFSS models and the observations near the Sun and at 1 astronomical unit (AU) do not fully match, often being only half or less of the heliospheric magnetic flux compared with in-situ observations \citep{Shi2024ApJglobal}. The open magnetic flux may not only originate from the observed dark regions in extreme ultraviolet (EUV) and X-ray wavelengths compared to observations. Moreover, it also indicates that the PFSS model is not sufficiently accurate \citep{rice2021global}.

To fix or improve the aforementioned problems, some modified global magnetic field models and MHD simulations are applied to obtain a more accurate global coronal model that includes the effects of the solar wind \citep{mackay2012models,wiegelmann2017coronalmagnetic,asvestari2024openfield}. For global magnetic field models, the spherical source surface assumption of the PFSS model is an over-simplified assumption. For different synoptic magnetic maps at solar maximums and minimums, the height of the source surface is not always around 2.5 $R_\odot$ \citep{boe2020TSE,Boe2024PFSS}, and the shape may not be spherical \citep{kruse2021pfssevaluation,huang2024adjustingpfss}. Some studies proposed ellipsoidal or more arbitrarily shaped source surfaces \citep{kruse2020pfsselliptic}, and have investigated how the height of the source surface changes over the solar cycle combining total solar eclipse images and in-situ observations \citep{huang2024adjustingpfss, Boe2024PFSS}. However, it is also noted that the variation of the source surface is difficult to predict and invert through observations. \citet{guo2012modeling} used the optimization method to relax the global and local PFSS model to a non-linear force free field, and it is shown that the model after relaxation has a more reliable magnetic field configuration. Moreover, some improved global magnetic field models have been proposed using magneto-frictional methods \citep{rice2021global,bhowmik2022exploring}, optimization methods \citep{mackay2005model, mackay2012optimation}, and other mathematical adjustments combined with MHD models to include solar wind effects \citep{yeates2018SSRv..214...99Y, Hayashi2021ApJMHDmf}. \citet{rice2021global} proposed an outflow field model that directly obtains the magnetic field in an equilibrium state of magneto-frictional relaxation, eliminating the need for time-consuming numerical extrapolation. It introduces a solar wind velocity profile into the magnetic friction speed, allowing the solved magnetic field to contain the influence of the solar wind. This model not only provides an innovative way to fix the open flux problem, but also can serve as a more computationally efficient initial condition of the magnetic field in global MHD simulations considering solar wind effects. 

For MHD simulations, there are many data-constrained and data-driven simulation software that incorporate the effects of the solar wind, emphasizing the necessity of considering impact of the solar wind on the distribution of coronal plasma and velocity \citep{Groth2000JGR...10525053G,Narechania2021solarwind,Lionello2023MAS, Hayashi2022ApJSteadystate,Feng2023MNRAS.519.6297F}. They can reproduce the properties of the solar wind with polytropic assumptions \citep{mikic1999mhd,keppens1999polytropic,perri2022coconut1} or complex heating mechanisms in the outer corona and the heliosphere space based on observed magnetograms or theoretical magnetic field models \citep{van2010awsom,antiochos2011solarwind,Usmanov2014threefluid, Usmanov2018SteadySW}. These codes can be divided into two categories by their simulation regions: the corona and the heliosphere. Some featured codes are as follows. Coolfluid Corona Unstructured (COCONUT) \citep{kuzma2023coconut,perri2022coconut1,perri2023coconut2} can generate a coronal model with streamers and an overall accurate velocity distribution, and its simulation domain ranges from 1 solar radius to 0.1 AU. The Magneohydrodynamic Algorithm outside a Sphere (MAS) code archives a time-dependent resistive thermodynamic MHD equations to simulate the coronal structures \citep{mikic1996large-scale,mikic1999mhd, linker2011openflux,mikic2018predict} and is included in the CORHEL (Corona-Heliosphere) model which can model the solar corona and inner heliosphere \citep{RILEY20121, torok2018sun2earth}. The Alfv\'en Wave Solar Model (AWSoM) is a global MHD modeling code under the Space Weather Modeling Framework (SWMF) using Alfv\'en wave turbulence to accelerate and heat the corona \citep{van2010awsom,Van2014AWSoM}. European heliospheric forecasting information asset \citep[EUHFORIA;][]{pomoell2018euhforia} is an MHD software that simulates the heliosphere from 0.1 AU to 2 AU. ICARUS \citep{verbeke2022icarus,baratashvili2022improving,baratashvili2024multi} is also a heliosphere model to simulate the propagation of the solar wind based on the Message Passing Interface Adaptive
Mesh Refinement Versatile Advection Code \citep[MPI-AMRVAC;][]{xia2018mpi,keppens2021mpi,rony2023mpi} framework, which allows for advanced adaptive mesh refinement (AMR) operations. However, most of the commonly-used corona models focus on a relatively large domain to $10\sim20~R_\odot$, where their top boundaries expand to super-Alfvénic and supersonic surface, the flow is dominant and the information is carried outward, without any feedback to the inner domain. Therefore, simple boundary conditions can be adopted at these outer boundaries. A near-Sun solar-wind-coupled corona model with its top boundary located in the sub-Alfvénic region can reveal finer interactions and features between the low corona and the ambient solar wind, while further requiring enhanced numerical stability to simulate the highly dynamic coupling between magnetic structures and the plasma flow field.

MPI-AMRVAC is widely used for simulations at the scale of active regions or prominence/filament structures, allows complicated heating mechanisms in the energy equation, and is both available for Cartesian and spherical coordinates. Therefore, it will be useful to implement a global corona model constrained by observed magnetograms using MPI-AMRVAC, taking into account the effects of the solar wind, basically for following reasons. First, such a solar corona model could provide a more realistic background magnetic field and using the same software garantuees a seemless interaction with local MHD simulations in spherical coordinates \citep{wyper2021model,wyper2024model} with possible finer AMR meshes. The possibility to (de)activate AMR on user-defined physical and geometric criteria will not only offer a more accurate velocity distribution but also provide a more precise magnetic field configuration, including large-scale structures such as streamers, pseudo-streamers and the open field. This then paves the way for follow-up simulations of magnetic flux rope (MFR) eruptions and magnetic topological evolution with interaction of solar-wind-affected structures \citep{fan2018mhd,fan2024data} up to higher radial altitudes. A near-Sun coronal model based on AMRVAC can directly benefit from the open-source development where continuously new physics and data-analysis modules are provided, can easily compare the effect of complex heating functions, and can connect to other heliosphere models including ICARUS, allowing for the study of CME propagation in the heliosphere \citep{linan2023self, linan2024cme, guo2024modeling}. Using the particle module of MPI-AMRVAC, we can directly extract local measurements that are comparable to in-situ observations, allowing to investigate the characteristics of CME propagation. A more direct application is the use of such a corona model in combination with observations from coronagraphs and TSEs to test the accuracy of this model \citep{badman2022constrain, wagner2022validation} with comparisons of the characteristics of streamers, filaments, and other large-scale features between simulations and observations. It could also be used to predict and diagnose the structure of the solar wind \citep{mikic2018predict,Liu2025arXivforecast}. 

In this work, we conduct a data-constrained MHD simulation \citep{guo2024review} of the global solar corona on 2024 April 8, when the 2024 TSE happened in the United States, with a reduced polytropic index $\gamma=1.05$ in the energy equation of ideal MHD equations. We focus on the magnetic field configuration and topology up to the region of high corona. The simulation domain spans from the photospheric boundary ($1~R_\odot$) to the outer corona ($2.5~R_\odot$), allowing us to study the dynamic effects of the solar wind within this region. Therefore, observations from TSE serve as critical validation benchmarks for the model. In Section~\ref{sec:observation}, we introduce the observations of the 2024 April 8 TSE in Texas and Arkansas of the United States in details of instruments, sites, and data pre-processing methods. In Section~\ref{sec:MHD}, we present the numerical schemes and MHD equations used in AMRVAC. Section~\ref{sec:mhd model} describes the MHD equations. We discuss the initial conditions of MHD simulations in detail in Section~\ref{sec:IC}, where we improve the outflow field model proposed by \citet{rice2021global}, compare the effects of different input magnetograms, and use it as the initial magnetic field. Section~\ref{sec:BC} introduces the boundary conditions we use for the MHD variables. In Section~\ref{sec:results}, we show results of our MHD simulations and compare them with TSE images. in terms of the magnetic field configuration in Section \ref{sec:results:magnetic field}, coronal variable distributions in Section \ref{sec:results:coronal variable}, and synthesize white-light radiation and pseudo-radiation from the current density in Section \ref{sec:results:radiation}. A summary and discussions on the advantages and future improvements of this corona model are presented in Section~\ref{sec:discussions}.

\section{Observations} 
\label{sec:observation}
The observations of the TSE event on 2024 April 8 were taken with two sets of instruments, one in the white-light waveband and the other the \ce{Fe XIV} emission line. The instruments are developed by the scientific expedition team of Nanjing University including Yuhao Huang, Sizhe Wu, Yihua Li, and Qinghui Lao under the guidance of Zhongquan Qu from Yunnan Observatories. The white-light images were obtained in Oden, Montgomery County, Arkansas, USA at N\ang{34;37;02}, W\ang{93;46;40}. The \ce{Fe XIV} images were obtained in Arlington, Tarrant County, Texas, USA at N\ang{32;43;46} and W\ang{97;07;33}. During the totality, Oden had clear skies, allowing full data collection for about $200~\mathrm{s}$, while Arlington experienced intermittent cumulus clouds, reducing usable data of about $120~\mathrm{s}$. 

The observational setup consisted of two imaging systems with identical imaging system, containing a ZWO AM5 mount coupled with an Askar FRA400 apochromatic refractor telescope featuring a $72~\mathrm{mm}$ aperture and $400~\mathrm{mm}$ focal length at f/5.6, along with a ZWO ASI294MM Pro camera equipped with $8288\times5644$ pixels and the size of the sensor is $19.1~\mathrm{mm}\times13.0~\mathrm{mm}$. The white-light imaging system employed one set of this basic configuration, while the \ce{Fe XIV} imaging system incorporated a dual-channel design with two sets of parallel telescopes and cameras, each fitted with narrow-band filters with the bandwidth at $1~\mathrm{nm}$ centered at $530.5~\mathrm{nm}$ of the Fe XIV emission line and $532.0~\mathrm{nm}$ at the continuum wavelength of this \ce{Fe XIV} line. This configuration achieves a field of view with $2.7\degree\times1.9\degree$, corresponding to $10.4~R_\odot\times 7.0 ~R_\odot$ with a scale of $1.2''$ per pixel, where one solar angular radius corresponds to $0.26\degree$. This field of view fully encompasses both the inner ($\sim 1.3R_\odot$) and middle corona ($1.3R_\odot\sim2.3R_\odot$) while partially capturing the outer corona regions ($> 2.3R_\odot$).

\begin{figure*}
    \centering
    \includegraphics[width=\textwidth]{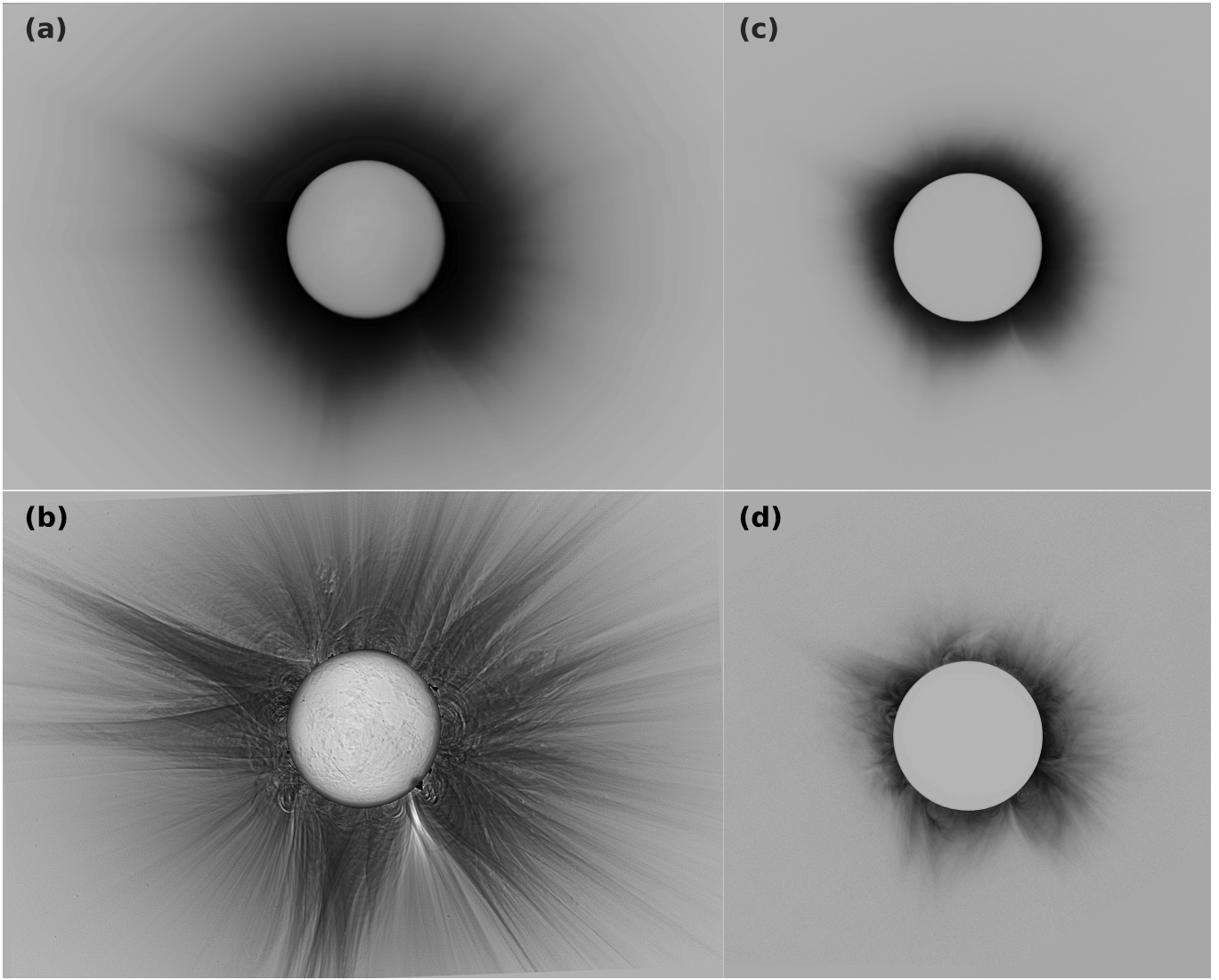}
    \caption{The images of 2024 April 8 TSE. (a) A typical original white-light wavelength image before the processing. (b) The processed white-light image after calibration, the Fourier transformation, the HDR, and the ACHF, with the moon disk displaying in its position. (c) The original image of \ce{Fe XIV} emission line at $530.5~\mathrm{nm}$. (d) The processed image of \ce{Fe XIV} emission line after subtracting the continuum waveband, the calibration, the Fourier transformation, and the ACHF.}
    \label{fig:observations}
\end{figure*}

During the TSE observation, the white-light images were acquired in 3 phases, namely the pre-totality, mid-totality and post-totality. Each phase contains images with 6 different exposure times with 20 images captured for each exposure time, and the last phase in the post-totality has only 5 exposure times due to the limited observation time. An example of the white-light raw images is shown in Figure \ref{fig:observations}(a). The \ce{Fe XIV} images are taken with a fixed exposure time of $0.03~\mathrm{s}$ and there are in total 86 images in the Fe XIV waveband $530.5~\mathrm{nm}$ and 116 images in the $532.0~\mathrm{nm}$ waveband. An example of the \ce{Fe XIV} images in the emission line waveband $530.5~\mathrm{nm}$ is shown in Figure \ref{fig:observations}(c). We develop a specialized image processing code for the TSE images based on the methods from \citet{2006CoSka..36..131D}, \citet{2009ApJ...706.1605D}, \citet{2014..8466..17} considering specific characteristics of this event. This code processes the observation data from raw images to the final coronal images with fine features.

For the white-light images in each phase, we process them with the following steps. First, a series of calibrated images are acquired with the calibration frames taken before the TSE happened and the raw data. Subsequently, we use the phase correlation method based on the Fourier transform to compute the transformation parameters between coronal images with identical exposure times. These images at the same exposure time level are then aligned through transformation and cropping. The equation of the phase correlation method is as follow:

\begin{equation}
n(x, y)=\mathcal{F}^{-1}\left(\frac{\mathcal{F}[a(x, y)] \cdot \mathcal{F}^{*}[b(x, y)]}{|\mathcal{F}[a(x, y)]| \cdot|\mathcal{F}[b(x, y)]|}\right)=\delta\left[\left(x-x_{0}\right) \bmod M,\left(y-y_{0}\right) \bmod N\right]\,,
\end{equation}
where $\mathcal{F},~\mathcal{F}^{-1}$ are the discrete Fourier transform and its inverse, $\mathcal{F}^*$ is the complex conjugate of $\mathcal{F}$, $ a(x,y) $ and $ b(x,y) $ are identical images related by a translation $ (x_0,y_0) $, and $M\times N$ is the dimension in pixels of the images. Theoretically, $ n(x,y) $ is an impulse function with its peak located at $ (x_0,y_0) $. The aligned images for each exposure time are then stacked to improve the signal-to-noise ratio. Next, the phase correlation method is used again to calculate the relative transformation parameters between these stacked master images of different exposure times. Transformation and cropping are applied to align these different exposure time images. After aligning these images using phase correlation, applying a mask to cover the lunar disk region, and detecting and removing stars on the background, a linear high dynamic range (HDR) combination based on exposure time is performed. The equation of the linear HDR combination is:

\begin{equation}
B=\displaystyle \frac{\sum_{i=1}^{n} \frac{A_{i}}{t_{i}} \cdot W_{i}}{\sum_{i=1}^{n} W_{i}}\,,
\end{equation}
where $ A_i $ represents the images with different exposure times, $ W_i $ is the corresponding weight, $ t_i $ is the exposure time, and $n$ is the number of images. This results in an HDR coronal image. Finally, the adaptive circular high-pass filter (ACHF) with different convolution kernel sizes, is applied separately in the tangential and radial directions to this HDR coronal image. The equation of the ACHF is:
\begin{equation}
B(r, \phi)=A(r, \phi)-A(r, \phi) * C(r, \phi)=1-\sum_{\rho=-k}^{k} \sum_{\theta=-k}^{k} C(\rho, \theta) \cdot A(r+\rho, \phi+\theta)\,,
\end{equation}
where $ A(r, \phi) $ is the HDR coronal image and $ C(r, \phi) $ is the one-dimensional Gaussian convolution kernel in either the tangential or radial direction. This generates a series of tangential and radial coronal images preserving different frequency structures. These images are synthesized separately within the tangential group and radial group, undergo wavelet sharpening, and finally, the sharpened tangential and radial images are combined to produce the final image with fine details, which is displayed in Figure \ref{fig:observations}(b). It is noted that there are distinguishable large-scale coronal structures, including the helmet streamers and coronal holes. There are also some small-scale structures, such as several loop systems, thin radial rays, polar plumes and discontinuities. Overall, this high-resolution white-light eclipse image provides a detailed, high-quality representation of coronal structures characteristic of solar maximum conditions. For the \ce{Fe XIV} images, the processing steps are identical to those for the white-light images, except that the linear HDR synthesis step is omitted. The final image of \ce{Fe XIV} is displayed in Figure \ref{fig:observations}(d). By comparison, it is seen that there are more fine features in the lower corona in \ce{Fe XIV}, including prominences, closed loops, and some plumes. 

\section{Numerical Setup and MHD Simulations} \label{sec:MHD}
\subsection{MHD model and Simulation Domain}
\label{sec:mhd model}
To investigate physical details consistent with observations, we perform an ideal data-constrained MHD simulation that includes an internal energy equation with a reduced polytropic index $\gamma=1.05$ in the spherical coordinates as supported by the open-source code MPI-AMRVAC 3.1. This software offers various options to numerically integrate the governing MHD set, and here we choose to work with a dimensionless and conservative form as follows:
\begin{gather}
    \label{mhd_rho}
    \frac{\partial \rho}{\partial t}+\nabla \cdot(\rho\mathbf{v})=0,\\
    \label{mhd_v}
    \frac{\partial (\rho\mathbf{v})}{\partial t}+\nabla\cdot(\rho\mathbf{vv}-\mathbf{BB}+\frac{\mathbf{B}^2}{2}\mathcal{I}+p\mathcal{I})=\rho\mathbf{g},\\
    \frac{\partial e_{\mathrm{int}}}{\partial t}+\nabla\cdot(\mathbf{v}e_{\mathrm{int}})=-p\nabla\cdot\mathbf{v},\\
    \frac{\partial\mathbf{B}}{\partial t}+\nabla\cdot(\mathbf{vB}-\mathbf{Bv})=0,\\
    \nabla\cdot\mathbf{B}=0,
\end{gather}
where $\rho$ is the density, $\mathbf{v}$ is the velocity, $p$ is the thermal pressure, $\mathbf{B}$ is the magnetic field, $\mathbf{g}=-g\mathbf{e}_r$ is the gravity acceleration, $e_{\mathrm{int}}=p/(\gamma-1)$ is the internal energy per unit volume, $\eta$ is resistivity, and $\mathbf{J}=\nabla\times\mathbf{B}$ is the electric current density. In this work, we take the constant adiabatic index to be $\gamma=1.05$, which is a reduction from the value $5/3$ usually adopted for a ideal gas with number of degrees of freedom $f=3$, and under this condition the simulated corona deviates from the actual physical conditions \citep{kuzma2023coconut}. However, this reduced adiabatic index can effectively approximate the effects of the solar wind without further considering complicated heating terms \citep{mikic1999mhd}. In future developments, we will adjust this reduced polytropic index approximation and account for more physical ingredients but here prioritize the coronal magnetic field topology and velocity distribution in the simulation.

The simulation domain is $\displaystyle [r_{\mathrm{min}},r_{\mathrm{max}}]\times[\theta_{\mathrm{min}},\theta_{\mathrm{max}}]\times[\phi_{\mathrm{min}},\phi_{\mathrm{max}}]=[1.001R_\odot,2.500R_\odot]\times[0\degree,180\degree]\times[0\degree, 360\degree]$. The domain is resolved by $400\times180\times360$ cells, and the grids along $r$-, $\theta$-, and $\phi$-directions are uniform. This resolution is significantly higher than that of other software, such as COCONUT, in the low coronal region, thus revealing more fine features, but it also requires more time to converge to a steady state. The Harten-Lax-van Leer (HLL) solver is used for flux computation, the minmod limiter is applied, and the generalized Lagrange multiplier (GLM) method \citep{dedner2002glm} is used for the divergence cleaning of the magnetic field. For the temporal integrator, we employ the Strong Stability Preserving Runge-Kutta 3rd order (SSPRK3) method \citep{Gottlieb1998TotalVD} for a three-step time integration \citep{rony2023mpi}, which satisfies the Total Variation Diminishing (TVD) condition and allows the Courant–Friedrichs–Lewy (CFL) condition to reach 1.

\subsection{Initial Conditions}
\label{sec:IC}
\subsubsection{The Outflow Field Model and Input Magnetograms}
\label{sec:IC:off}
The outflow field model proposed by \citet{rice2021global} inserts a radial solar wind velocity profile into the magneto-frictional method, and it can directly obtain the global magnetic field in an equilibrium state of the magneto-frictional relaxation with a pure radial solar wind through specific mathematical derivations and spherical harmonic expansions, which is derived with Eqs. (6), (7), and (9--13) in \citet{rice2021global}. Compared to the PFSS model, it provides different open fluxes by adjusting on the solar wind velocity, thereby allowing flexible adaptation based on different solar activity and specialties of events. In this work, we improve this model with a more precise Parker solar wind profile solved with Lambert functions \citep{Cranmer2004lambertPKSW}, which is specified as follows:
\begin{equation}
\label{eq:parker's solar wind solution}
\begin{aligned}
    &\dfrac{v_\mathrm{out}}{c_s}=
    \begin{cases}
        &\sqrt{-W_0[-D(r)]}, \quad r/r_c\le 1\\
        &\sqrt{-W_{-1}[-D(r)]}, \quad r/r_c\ge 1
    \end{cases},\\
    &D(r)=\left(\dfrac{r}{r_c}\right)^{-4}\exp{\left[4\left(1-\dfrac{r_c}{r}\right)-1\right]}.
\end{aligned}
\end{equation}
In Eq.~(\ref{eq:parker's solar wind solution}), $r$ is the radius, $v_\mathrm{out}(r)$ is the radial velocity of the Parker solar wind, $c_s=\sqrt{\gamma k_bT_0/\mu m_p}$ is the isothermal sound speed which is determined only by the isothermal temperature of the initial atmosphere $T_0$ and the average molecular weight $\mu=0.6$ for the mixture of hydrogen and helium, $r_c=\sqrt{GM/2c_s^2}$ is the critical radius of the sonic point, and $W_0$ and $W_{-1}$ are two branches of Lambert $W$ function. This modification gives an analytical solution to the Parker solar wind instead of an approximation in Eq.~(7) in \citet{rice2021global}. It can also derive a more accurate expression of the critical radius $r_c$ and remove the upper limit on the maximum solar wind velocity attainable at the source surface $v_1$, which is $v_1\sim 150~\mathrm{km/s}$ in the original model in \citep{rice2021global}, thereby extending the applicability of the outflow field model to solar maximum conditions. It also allows to adjust the top boundary of the outflow field model freely. Moreover, since we are using a uniform, non-staggered mesh in spherical coordinates, the boundary conditions we employ also differ from \citet{rice2021global}. With the use of the outflow field model as the magnetic field initial condition, the MHD relaxation time can be effectively diminished, which accelerates the convergence of the simulated corona with solar wind effects toward a stable state. The typical temperature or the solar wind velocity of the initial conditions are adjustable with the solar-cycle dependency.

The code we use to calculate the outflow field is MagWind\citep{magwind2025}, which is developed by authors of this article in Python and aimed for modeling the solar magnetic field enabling CPU/GPU parallel computing, available on Github\footnote{https://github.com/lavenderLi09/MagWind}. It supports the calculation of the PFSS, outflow field, and Schatten Current Sheet (SCS) models using both the theoretical and observed global magnetograms on stretched and uniform grids, which can serve as the initial magnetic field inputs of AMRVAC or other softwares. The SCS model is not currently employed in this study, however, it presents a valuable framework for future research extensions, particularly when investigating a larger domain. 

For the bottom boundary of the outflow field model, we compare different synoptic maps and the daily update synchronic frames from the data series of the Solar Dynamics Observatory/Helioseismic and Magnetic Imager \citep[SDO/HMI;][]{2012SoPhHMI}, Global Oscillation Network Group \citep[GONG;][]{1996SciGONG}, and Air Force Data Assimilative Photospheric Flux transport \citep[GONG-ADAPT;][]{2010AIPCAPADT}. For the synoptic maps, The HMI synoptic map provides the radial component of the magnetic field over an entire Carrington rotation with a higher spatial resolution of $3600\times1440$ compared to magnetograms from GONG and GONG-ADAPT \citep{perri2023coconut2}. It is assembled with pseudo-radial magnetic field data at different dates, instead of simultaneous observations. Therefore, the synoptic map exhibits temporal displacement. The data at any given Carrington longitude represent an average of observations taken as that longitude rotated across the solar central meridian\footnote{http://jsoc.stanford.edu/HMI/LOS\_Synoptic\_charts.html}. For example, for the date corresponding to the 0° longitude on the map, the data on the synoptic map to its right (increasing longitude) was acquired in the preceding days, while data to its left (decreasing longitude) was acquired approximately one Carrington rotation earlier ($\sim27$ days prior). Consequently, the synoptic map does not provide a snapshot-like accuracy of the true photospheric radial magnetic field distribution on any single specific day during the Carrington rotation, especially for dates close to the 0° or 360° longitude. This temporal displacement effect became particularly significant during this TSE event on 2024 April 8. HMI synoptic map of Carrington rotation 2282 ends on 2024 April 9, and therefore can not provide a quantitatively reliable approximation of the radial magnetic field $B_r$ distribution of this TSE.

\begin{figure*}
    \centering
    \includegraphics[width=\textwidth]{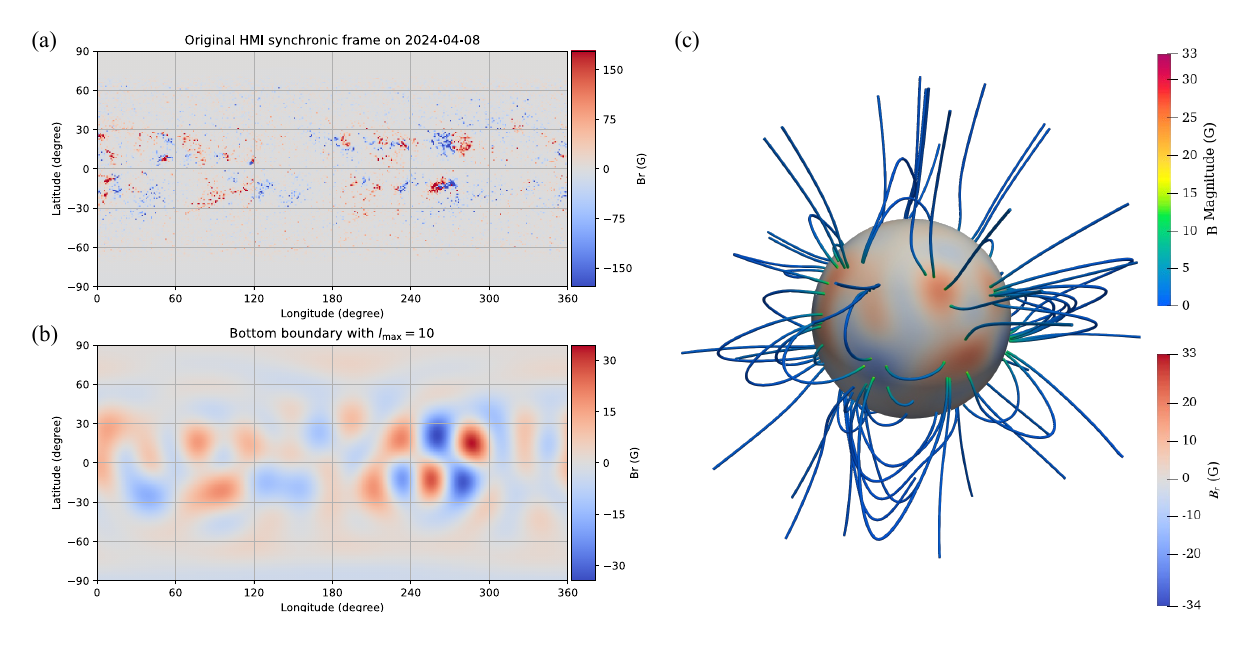}
    \caption{The HMI synchronic frame, bottom boundary of the outflow field, and the outflow field model. (a) The HMI synchronic frame on 2024 April 8 with a resampled resolution $360\times180$ and axis (longitude-latitude). The ranges of the colorbar have been set to plus and minus of the maximum of the field intensity divided by 10 in order to display a more detailed magnetic field polarity distribution. (b) Bottom boundary of the outflow field model with $l_{max}=10$. The ranges of the colorbar display the original field intensity. In both (a) and (b) the positive polarities are in red and negative polarities in blue. (c) Magnetic field configuration of the outflow field model with the bottom boundary displaying the $B_r$ distribution on $r=1.002~R_\odot$ plane. }
    \label{fig:off}
\end{figure*}

The synchronic frame replaces the first 120° in longitude from the original synoptic map with the daily full-disk magnetic field data at the observed date. In this way that part of the synoptic frame better represents the field observed simultaneously with more up-to-date data. We have compared the synchronic frames from SDO/HMI, GONG and GONG-ADAPT, calculated the PFSS and outflow field models based on them with MagWind, and compared the positions and shapes of the magnetic field loops to the streamers from TSE observations. We find that HMI magnetograms offer higher spatial resolution and employ a more accurate polar field filling method, resulting in reconstructed magnetic features that show a closer similarity to observations. Therefore, we use the HMI synchronic frame map from the series \texttt{hmi.mrdailysynframe\_polfil\_720s} on 2024 April 8 as the bottom boundary of the initial magnetic field in Figure \ref{fig:off}a and the central median of this day is at 60° in longitude from the left-leading edge. An outflow field with the spherical harmonic order $l_{max}=10$, the source surface $R_{ss}=2.5R_\odot$ and $v_1=40~\mathrm{km/s}$ is used as the initial magnetic field after multiple parameter tests, and is displayed in Figure \ref{fig:off}c, and its bottom boundary is shown in Figure \ref{fig:off}b. It is noted that the overall magnetic field strength has been smoothed with $l_{max}=10$, therefore exhibiting its maximal is $\sim30~\mathrm{G}$. This is a scale filtering procedure which achieves an optimal balance considering critical smaller structures including solar active regions while excluding excessively fine details that would slow down the computational performance \citep{perri2023coconut2}. The Outflow field model computed in Magwind is on a full global spherical grid, and is then used as the input of the AMRVAC mesh. 

\subsubsection{The Solar Wind Profile}
\label{sec:IC:other variables}
For initial conditions of the coronal atmosphere, we initialize it by 1D Parker solar wind solution. The transcendental equation of the Parker solar wind is:
\begin{gather}
    \label{eq:parker sol}
    \frac{v_{out}^2}{c_s^2}\mathrm{exp}(1-\frac{v_{out}^2}{c_s^2})=\frac{r_c^4}{r^4}(4-4\frac{r_c}{r}),
\end{gather}
and it is solved also with the Lambert functions in Eq.~(\ref{eq:parker's solar wind solution}). The initial density is obtained from conservation of density flux, namely $r^2\rho v=const.$, based on the solar wind velocity with the density on the bottom boundary $\rho_b=1.17\times10^{-16}~\mathrm{g\cdot cm^{-3}}$ in the physical domain. In the ghost cells we assume that the density is constant at $\rho_b$. Therefore, the initial density profile is radially structured, aligned with the initial outflow magnetic field.

\begin{figure*}
    \centering
    \includegraphics[width=\textwidth]{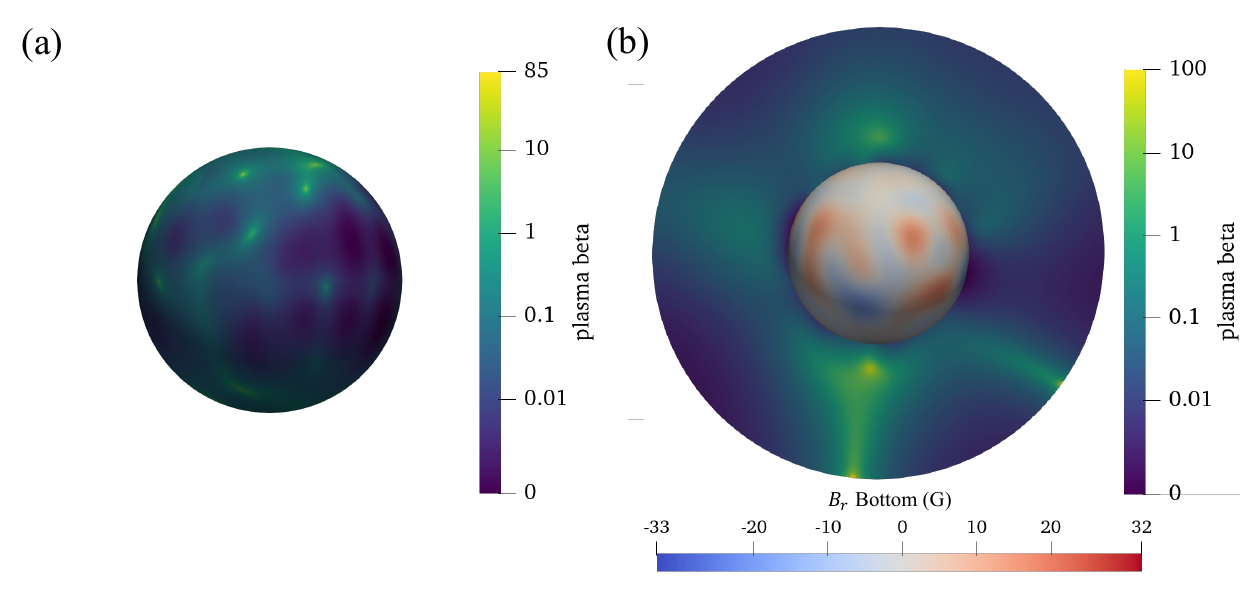}
    \caption{(a) Plasma $\beta$ distribution on the bottom surface and (b) the meridian plane.}
    \label{fig:beta}
\end{figure*}

The initial thermal pressure is constrained by an isothermal temperature $T_0=2.0\times10^{6}~\mathrm{K}$. A series of tests were performed to obtain an initial density distribution that more closely aligns with realistic physical conditions, while is compatible with the strength of the initial magnetic field intensity, so that the overall plasma beta is within a reasonable range shown in Figure \ref{fig:beta}, similar to other similar studies \citep{brchnelova2023A&Abeta,Cai2025MHDmodeling}. On the bottom surface, the minimum of the plasma $\beta$ is $\sim 0.001$ and the maximum of the plasma $\beta$ is $\sim 10^1$ on the positions above closed field and higher corona, which are within a reasonable range under the situation of the reduced polytropic index $\gamma=1.05$. 

\subsection{Boundary Conditions}
\label{sec:BC}
At the bottom boundary where $r=1.001R_\odot$, the density, the radial velocity, and thermal pressure on the two ghost layers are fixed at their initial values obtained from the Parker solar wind solution as line-tied conditions. The tangential velocity components are set to zero. Therefore, these cells serve as a reservoir of the coronal plasma to sustain the wind solutions \citep{wyper2021model}. For the magnetic field, all components $B_r,~B_t,$ and $B_p$ are fixed to the initial outflow field model in the inner ghost layer closest to the physical domain, while for the outer ghost layer, a third-order equal-gradient extrapolation is employed. At the upper boundary where $r=2.5R_\odot$, a zero-gradient extrapolation method is applied for the density, thermal pressure and radial velocity. Moreover, in order to better maintain the divergence-free magnetic field $\nabla\cdot\mathbf{B}=0$, we use a zero-gradient extrapolation for $r^2B_r$ at the outer radial boundary. At the outer radial boundary, all the tangential components for the velocity and magnetic field $v_t,~v_p,~B_t,$ and $B_p$ are set to zero, which are referred as half-slip conditions. The equal-gradient extrapolation means the derivatives of two adjacent nodes are equal, namely $u_i'=u_{i+1}'$, expressed as $u_i=\frac{29}{11}u_{i+1}-\frac{27}{11}u_{i+2}+u_{i+3}-\frac{2}{11}u_{i+4}$ in the one-sided finite difference scheme and the third-order accuracy, where $u$ is the corresponding physical quantity. The zero-gradient extrapolation used on the top boundary requires the derivative $u_i'=0$ expressed as $u_i=\frac{4}{3}u_{i+1}-\frac{1}{3}u_{i+2}$ in the one-sided finite difference scheme and the second-order accuracy.

Aforementioned boundary conditions allow for capturing dynamical evolution within the near-Sun global corona, which lies in largely subsonic and sub-Alfvénic regions. This allows the plasma coupling with magnetic field to exhibit more complicated dynamic and evolution patterns, as discussed in previous similar studies \citep{Hayashi2023ApJSinner}.

At the geometric singularities along the spherical polar axis ($\theta=0,~\pi$), we apply a pole boundary treatment that combines symmetric and antisymmetric conditions, enabling a smooth passage of mass and momentum across the polar regions. The ghost cells of $(r,\theta,\phi)$ at pole boundaries are filled by copying values from diametrically opposite grid cells at coordinates $(r,\theta,\phi+\pi)$. This $\pi$-periodic pairing identifies the source cells across the pole, effectively transforming the physical pole boundary into an internal computational boundary. Specifically, the scalars and radial components of vectors $B_r,~v_r$ adopt symmetric conditions, while the transverse components $B_\theta,~B_\phi,~v_\theta,~v_\phi$ adopt antisymmetric conditions, which ensure the continuity and stability at the singularities. For the $\phi$ boundaries at $\phi=0$ and $\phi=2\pi$, we adopt periodic conditions \citep{vanrony2007hybridAMR}. All the aforementioned boundary conditions presented are the final choice after rigorous testing, ensuring both computational stability at the boundaries and a physically faithful representation of the solar wind atmosphere.

\section{Simulation Results}
\label{sec:results}
\begin{figure*}
    \centering
    \includegraphics[width=\textwidth]{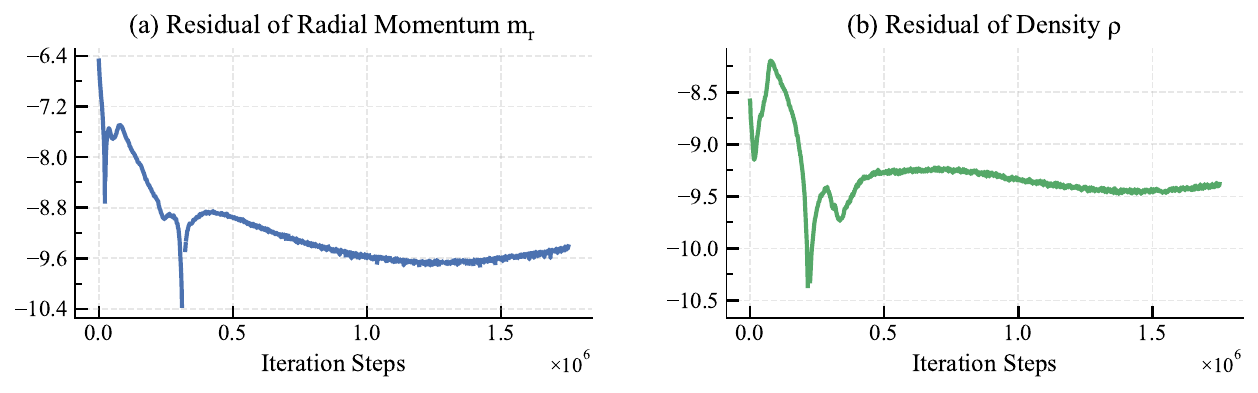}
    \caption{The evolution history of the residual of the radial-component of momentum $m_r$ and density $\rho$. The number on the x-axis is the iteration steps, and the y-axis represents the residual of $m_r$ and $\rho$ in the base-$\mathrm{10}$ logarithm.}
    \label{fig:02evolution}
\end{figure*}

We successfully conduct an MHD relaxation with solar wind effects of 2024 April 8 TSE event. Figure \ref{fig:02evolution} shows the evolution history of this simulation. The residual is defined similarly to \citet{kuzma2023coconut} as below:
\begin{equation}
    \mathrm{res}(a)=\mathrm{log_{10}}\sqrt{(a^t-a^{t+1})^2},
\end{equation}
where $a^t$ is the average value over the computation domain of a certain variable, e.g., momentum components, density and/or pressure, at the $t^{th}$ time step. After a period of relaxation, the magnitude of residuals stabilizes and no longer exhibits significant changes with increasing simulation time, indicating that we have obtained a steady-state solution with minimal variation over time steps. This suggests that the solar wind atmosphere has relaxed into an almost stable state. Figures \ref{fig:02evolution}a and \ref{fig:02evolution}b  illustrate the evolution of residuals of the radial-component of momentum $m_r$ and density $\rho$, respectively. It is shown that after $1.5\times10^6$ CFL limited iteration steps of relaxation, they all reach local minimums with negligible further reduction. Our subsequent simulation analyses were conducted thereafter. In this work, the relaxation was performed using 960 CPUs and was completed in 400 minutes physical time, taking 72 hours of computational time. 

The evolution of the open magnetic flux of the MHD relaxation was examined and compared with that of the initial Outflow field and a PFSS model based on the same HMI synchronic frame. The PFSS model yields an open flux of $\sim 2.23\times10^{22}$ Mx, the Outflow field $\sim 3.51\times10^{22}$ Mx, and the MHD-relaxed model $\sim 3.88\times10^{22}$ Mx. These results indicate that the prescribed solar-wind profile in the Outflow field model effectively increases the open flux compared with the PFSS, and that the subsequent MHD relaxation further increases it, driving the corona toward a quasi-steady state with a higher mean outflow velocity and locally inhomogeneous structures related to magnetic topology.

\subsection{Simulated Magnetic Field Configuration and Topology}
\label{sec:results:magnetic field}
\begin{figure*}
    \centering
    \includegraphics[width=\textwidth]{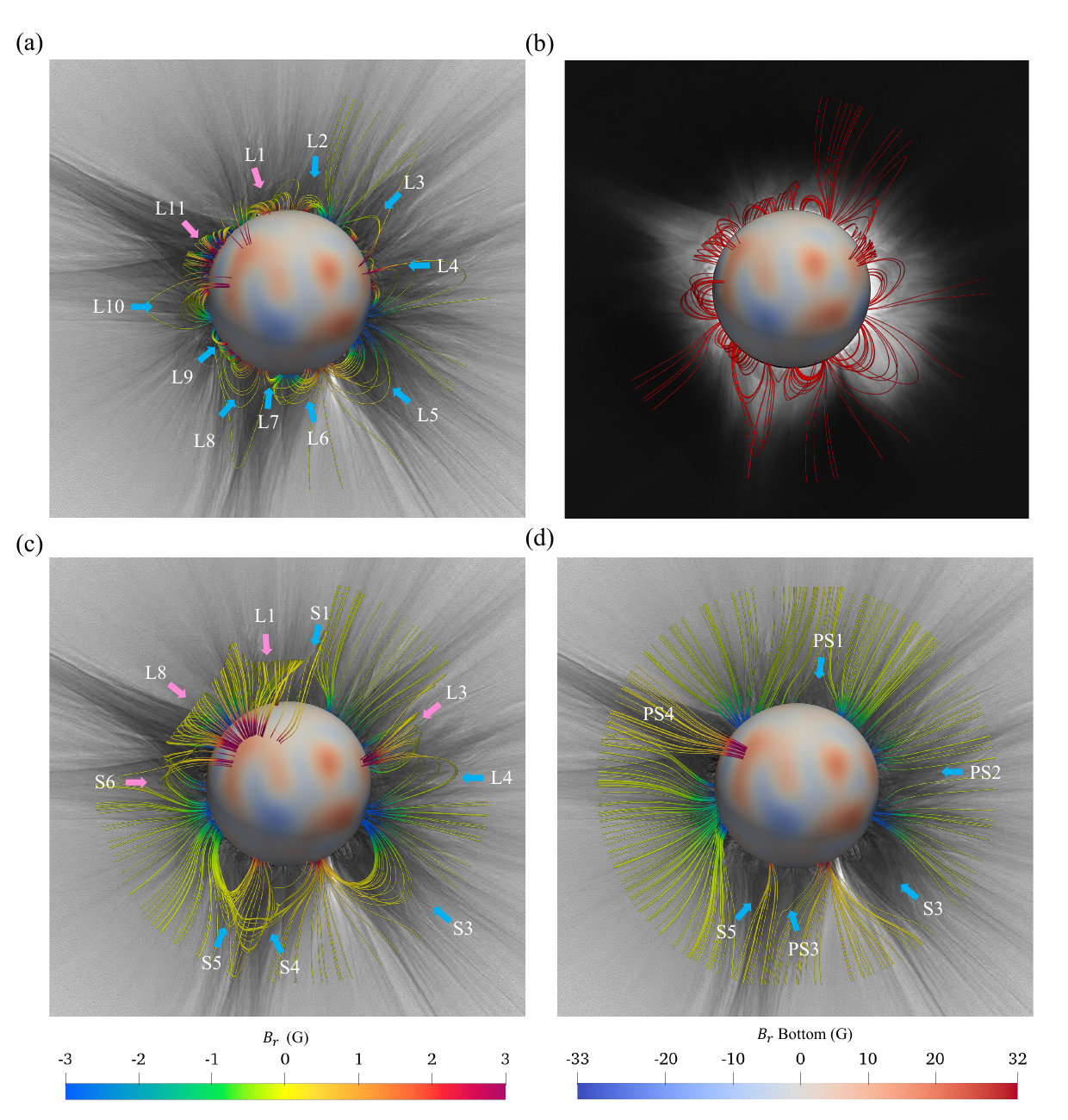}
    \caption{The solar coronal magnetic field lines from the MHD simulations on 2024 April 8 overlaid with TSE observations. The footpoints of the magnetic field lines are selected on the meridian plane orthogonal to the Sun-Earth axis. (a) The magnetic field lines traced from a line at $r=1.05~R_\odot$ on the meridian plane, overlaid with the white-light image of this TSE, and the bottom plane shows the distribution of $B_r$ at $r=1.005~R_\odot$. The labeled features correspond to eleven closed magnetic field loops in the lower corona. The arrows indicate the positions of magnetic loop, and blue arrows denote loops whose axes are essentially parallel to the meridian plane, while pink arrows represent loops whose axes are perpendicular to the meridian plane. (b) Field lines traced similar to (a), overlaid with the emission from \ce{Fe XIV} waveband. (c) Field lines traced from a line at $r=1.5~R_\odot$ on the meridian plane, overlaid with the white light image. The labeled features correspond to four loops and five streamers, and their numbers are consistent with those shown in (a). The arrows and their colors maintain the same definitions as in (a). (d) Field lines traced from a line at $r=2.3~R_\odot$, and others same as (c). The labeled features correspond to four pseudo-streamers and two streamers traced from relatively high altitudes. The numbers of labels, the arrows and their colors are consistent with (a).}
    \label{fig:03maglines}
\end{figure*}

We compare the steady-state simulated magnetic field configuration in Figure \ref{fig:03maglines} with the TSE observations in white light and the emission line from \ce{Fe XIV} in Figure \ref{fig:observations}b and \ref{fig:observations}d, respectively. Large-scale structures including streamers are clearly visible in white light wavelength, while closed magnetic field loops are more extinct in the \ce{Fe XIV} emission. Therefore, white-light images are more suitable for comparison with higher, larger-scale magnetic field features in simulations, while Fe XIV is more suitable for contrasting with lower, smaller-scale closed field.

By comparing the magnetic field in Figure \ref{fig:03maglines} and Figure \ref{fig:off}c, it is noted that after the MHD relaxation to a steady state from its initial original topology as shown in Figure \ref{fig:off}c, the overall magnetic field configuration undergoes a comprehensive transformation: new closed fields emerge, additional open fields form through reconnection, and the cusps of closed fields become significantly sharper, which aligns better with characteristic streamer morphologies affected by solar wind dynamics. Figures \ref{fig:03maglines}a and \ref{fig:03maglines}b indicate that the positions and morphologies of most small-scale closed fields traced from $r=1.05~R_\odot$ show a good match with observations, with the bottom plane displaying $B_r$ at $r=1.005~R_\odot$, and the color mapping of the magnetic field lines quantitatively represents the strength of the radial magnetic field $B_r$. There are ten labeled closed magnetic field loops in Figures \ref{fig:03maglines}a and \ref{fig:03maglines}b in total. The blue and pink arrows in Figure \ref{fig:03maglines}a denote different orientations of the loop axes, with the blue arrows indicating axes parallel to the meridian plane and the pink arrows of L1 and L11 representing axes perpendicular to the plane. From the background observations of the white light and \ce{Fe XIV} emission, it is noted that the positions and shapes of the simulated loops in the low corona show good agreement with observations, which demonstrates the accuracy of the simulation. Figures \ref{fig:03maglines}c and \ref{fig:03maglines}d display the magnetic field configuration from a higher plane at $r=1.5~R_\odot$ and $r=2.3~R_\odot$, respectively. In Figure \ref{fig:03maglines}c, certain closed loops, such as L1, L3, and L8, expand and stretch upward, but are still closed field. The loops L3 and L10 in Figure \ref{fig:03maglines}a become sharper with the orientation of its axis becomes vertical, namely L3 and S6 in Figure \ref{fig:03maglines}c. On the plane of $r=1.5~R_\odot$, several streamers have evolved, exemplified as S1, S3, S4, and S5. Similarly, it can be found that the magnetic field configuration on the west limb is in good agreement with observations, such as the open field between S1 and L3, the morphology and locations of L4 and S3. Moreover, the streamer-like magnetic field features including S1 at the north pole and S4, S5 at the south pole correspond well with the TSE observations. In Figure \ref{fig:03maglines}d, there are mainly four pseudo-streamers PS1--4, two streamers S3 and S5, and the open field. Within the simulation domain, both the height, shape, and positions of the closed magnetic field loops, the streamers, the pseudo-streamers, and the open field are similar to observations, which indicates the reliability of the MHD simulations. It is also noted that the magnetic field lines traced from higher surfaces in Figure \ref{fig:03maglines}d reveal that some open field lines, which envelop the outer sides of low-lying loops and streamer structures, evolve into a pseudo-streamer configuration.

However, the magnetic field near the north direction on the east limb failed to reproduce the observed features accurately, exhibiting some loops perpendicular to the meridional plane, such as L11 in Figure \ref{fig:03maglines}a, and L8, S6 in Figure \ref{fig:03maglines}c. This discrepancy appears in many simulations of this TSE \citep{Liu2025arXivforecast, cooper2025science}, and similar simulations for solar maximum periods have also identified similar drawbacks \citep{kuzma2023coconut}. This is mainly due to the following reasons. The HMI synchronic frames we used only replace the longitudinal range within $\pm 60\degree$ of the central meridian in synoptic maps with daily observed magnetic field data, which results in discrepancies between the magnetic field data used and the actual global magnetic field on the TSE day, which spans at least ±90° longitude and may be affected by the magnetic field features within a wider longitude range. Actually, comparison with the synchronic frame from 2024 April 15, namely, one week after the eclipse, reveals that some strong active regions had emerged on the solar west limb \citep{cooper2025science}. Generally, this TSE event happened within the solar maximum, and the solar activity is intense, resulting in more dynamic and transient magnetic field variations. As mentioned in Section \ref{sec:IC:off}, the observed magnetograms fail to fully capture the accurate magnetic field information, consequently adding constraints on simulations.

\begin{figure*}
    \centering
    \includegraphics[width=\textwidth]{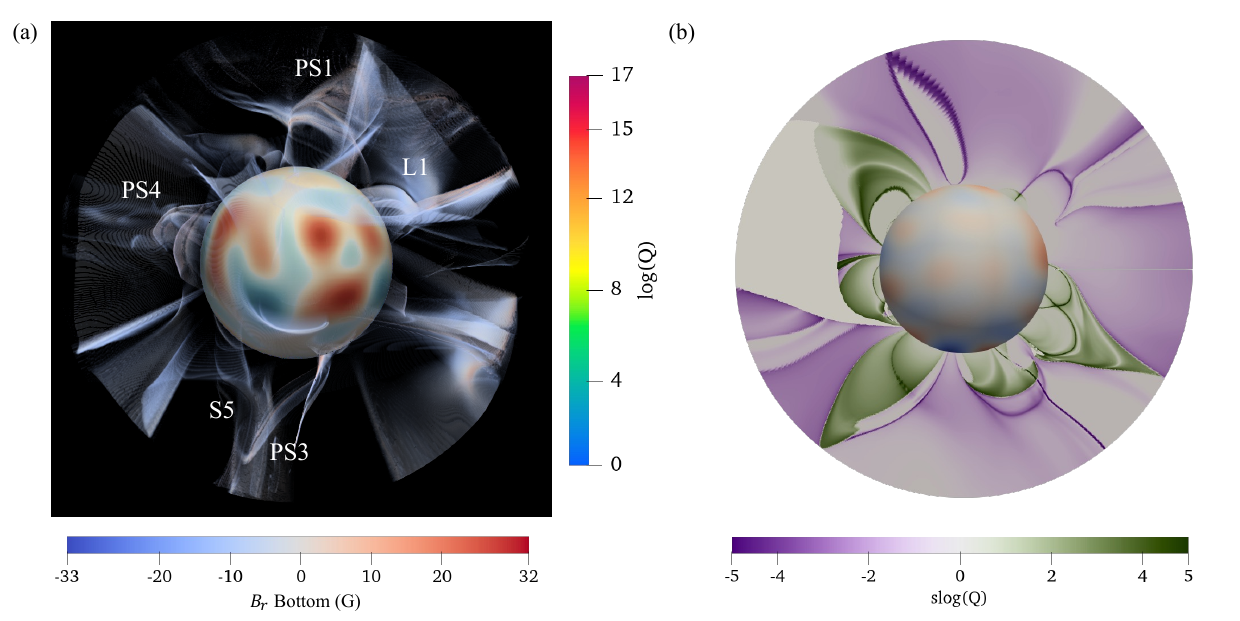}
    \caption{The squashing factor $Q$ in volume-rendered fashion and its distribution on a selected plane. (a) The volume-rendered distribution of $\mathrm{log}_{10}(Q)$ with labeled features corresponding to the magnetic field structures in Figure \ref{fig:03maglines}. (b) The distribution of $\mathrm{slog}(Q)$ along the small circle at $\theta=144\degree$. }
    \label{fig:07QSL}
\end{figure*}

Moreover, we also calculate the quasi-separatrix layers (QSLs) of this magnetic field configuration, and the results are presented in Figure \ref{fig:07QSL}. The calculation is accomplished by FastQSL \citep{fastqsl2021}, which is an advanced QSL calculation software that can be efficiently and accurately applied to simulation data in spherical coordinates. From Figure \ref{fig:07QSL}a, it is noted that there are distinguished features corresponding to the magnetic field structures including the pseudo-streamers PS1, PS3, and PS4, loop L1, and streamer S5. The positions with relatively high Q-values generally align to locations of closed loops and cusps. We also calculate the signed squashing factor $\mathrm{slog}(Q)=\mathrm{sign}(Q)\mathrm{log}(|Q|)$, which is commonly used to diagnose the magnetic field topology in Q-maps and S-Web \citep{Titov2011ApJtopology, Aslanyan2024fieldline}. The distribution of $\mathrm{slog}(Q)$ in the $\theta=144\degree$ plane is shown in Figure \ref{fig:07QSL}b, and it represents a typical region with strong magnetic field distribution, from which the relaxed helmet streamers can be clearly identified with high and positive $\mathrm{slog}(Q)$, and the open field with negative $\mathrm{slog}(Q)$. The relaxed streamers exhibit distinctly sharp cusps and have complex magnetic field features inside. Overall, our simulation successfully captures the majority of large-scale helmet streamers, demonstrating the validity of this MHD simulation.

\subsection{Coronal Variable Distribution}
\label{sec:results:coronal variable}
\begin{figure*}
    \centering
    \includegraphics[width=\textwidth]{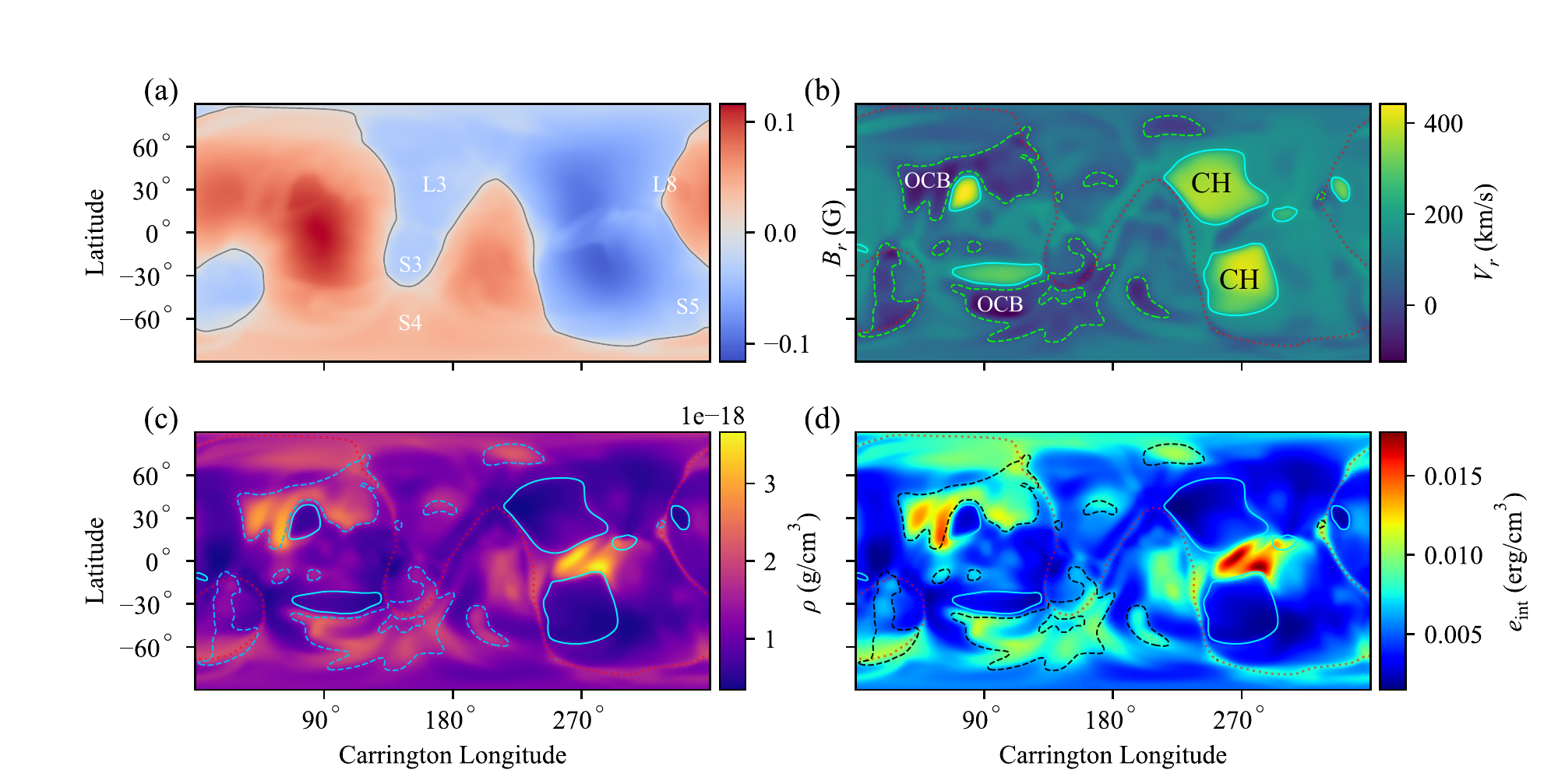}
    \caption{Synoptic maps of the radial magnetic field intensity (a), the radial velocity (b), the density (c), and the internal energy (d) at $2.49~R_\odot$. The gray line in (a) denotes polarity inversion line of $B_r$, and the labeled features L3, L8, S3, S4, and S5 correspond to the featured magnetic field structures in Figure \ref{fig:03maglines}. The red lines in (b), (c) and (d) denote the same $B_r$ PIL. The labeled features CHs and OCBs in (b) mark distinct coronal hole regions with high radial velocities, and the boundaries between open-field and closed-field regions with slow velocities.}
    \label{fig:04SS}
\end{figure*}

We analyze here also the thermodynamic and flow quantities derived from the MHD simulation. Figure \ref{fig:04SS} displays the synoptic maps of this MHD simulation at $r=2.49~R_\odot$, close to the source surface. Figure \ref{fig:04SS} presents distributions of the radial magnetic field $B_r$, the radial velocity $v_r$, the density $\rho$, and the internal energy $e_{\mathrm{int}}$. The polarity inversion line of $B_r$ is also displayed in Figure \ref{fig:04SS}a with a gray line, which indicates the position of the heliospheric current sheet (HCS). The distributions of density in Figure \ref{fig:04SS}c and radial velocity in Figure \ref{fig:04SS}b reveal that the HCS exhibits spatial coincidence with relatively high density, high internal energy, and low radial velocity areas. These results are consistent with previous studies \citep{Feng2015Data, wang2025ApJSsip}. In Figures \ref{fig:04SS}b, \ref{fig:04SS}c, and \ref{fig:04SS}d, two coronal hole (CH) regions are identified, characterized by low density and high radial velocity plasma flows along open field lines originating from areas exhibiting strong photospheric magnetic fields. Several slow-wind regions near open – closed field boundaries (OCBs), located around streamers and adjacent open field boundaries, are also marked. The loops L3, L8 and streamers S3, S4, and S5 in Figure \ref{fig:04SS}a generally correspond to structures in Figure \ref{fig:03maglines}, displaying relatively lower velocity, higher density, and enhanced internal energy near the source surface, occasionally influenced by nearby open fields (e.g., L8 here and PS4 in Figures \ref{fig:03maglines}d). These distributions of $B_r$, $\rho$, $v_r$, and $e_{\mathrm{int}}$ near the top boundary support the physical consistency of the model, with the high numerical resolution enabling the capture of fine and dynamic features.

\begin{figure*}
    \centering
    \includegraphics[width=\textwidth]{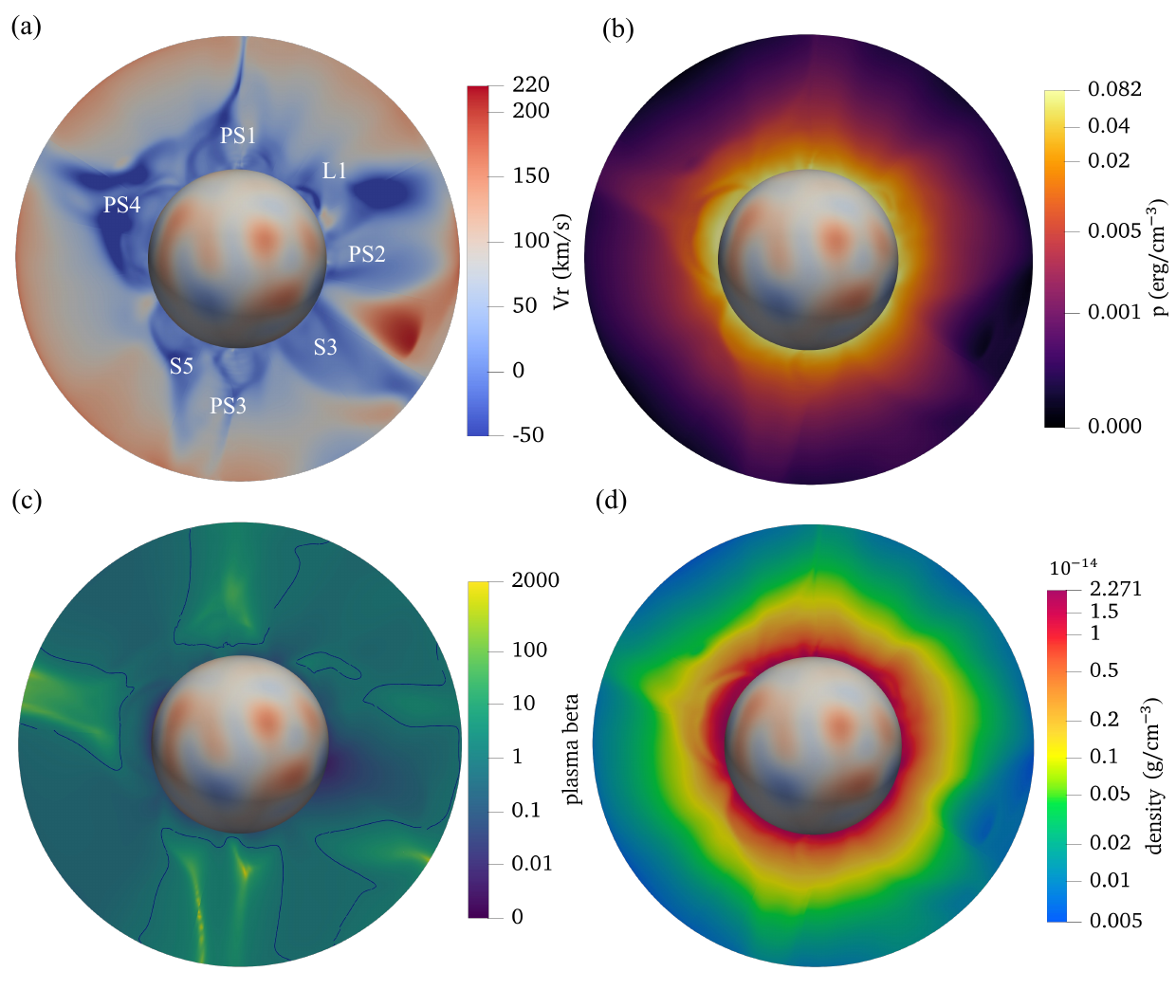}
    \caption{The meridian plane displays (a) the radial velocity, (b) the thermal pressure, (c) the plasma beta, and (d) the density of MHD simulations with bottom boundary displaying the radial magnetic field strength at $r=1.005~R_\odot$. The blue contours in (c) denote the position where $\beta=1$. The labeled structures in (a) denote four pseudo-streamers, two streamers, and one loop, which are clearly distinguishable in the velocity distribution, and the labeled features are consistent with Figure \ref{fig:03maglines}a.}
    \label{fig:05variable}
\end{figure*}

We also examine the features of the overall coronal magnetic field including streamers with the distributions of the radial velocity $v_r$, the thermal pressure $p$, the plasma beta $\beta=p/(\mathbf{B^2}/2\mu_0)$, and the density $\rho$ on the meridian plane on the day of the TSE in Figure \ref{fig:05variable}. From the radial velocity distribution in Figure \ref{fig:05variable}a, it is noted that there are clearly multiple regions exhibiting the characteristics of streamers with significantly low velocity and cusp structures. We mark these regions respectively into three categories, which are streamers, pseudo-streamers, and closed loops. It is noted that regions beneath loops and helmet streamers exhibit negative velocities, and there are two primary reasons for the formation of this pattern. One is that magnetic reconnection above the streamer generates both upward and downward outflows, resulting in negative velocities below the reconnection site. The other is that the strong magnetic field near the lower boundary may trigger complex dynamical behavior in this region during the MHD evolution. Such behavior becomes even more evident in our high-resolution MHD simulations focusing on the near-Sun region.

There are four distinct pseudo-streamers in the velocity distribution in Figure \ref{fig:05variable}a in total, which are relatively narrow, confined, and have rather complicated structures in the base regions, such as double-lobed or multi-lobed configuration. PS1 and PS3 are in the polar regions, and they correspond to the magnetic field structures PS1, PS3 in Figure \ref{fig:03maglines}d. PS2 is located at the central west limb of the solar disk, exhibiting no distinct cusp structure but maintaining correspondence with PS2 in Figure \ref{fig:03maglines}d. It displays helmet-like magnetic field line configurations that spatially align with loop L4 in Figures \ref{fig:03maglines}a and \ref{fig:03maglines}c. However, it should be noted that the top of PS2 does not exhibit a sharp, typical streamer structure, and the overall orientation of the feature is tilted further southward compared to the TSE observation. This is possibly because of the limited simulation domain and the magnetic field in this region did not fully relax in the MHD simulation. PS4 is situated in the northeastern direction, exhibiting an extended morphology with complex structural features. These pseudo-streamers all contain low-velocity plasma bulks. These typical pseudo-streamers are also consistent with the TSE observations, which exhibit several pseudo-streamers in the south pole, the north pole, and the northeastern direction. Moreover, there are also two streamer-like low-velocity regions, S3 and S5, distributed close to the south polar region. They correspond to the magnetic field structures S3 and S5 in Figures \ref{fig:03maglines}c and \ref{fig:03maglines}d, and are related to the closed loops L5 and L8 in Figure \ref{fig:03maglines}a. L1 is a distinctly visible low-velocity and compact loop that matches the L3 magnetic field loop in Figure \ref{fig:03maglines}a. Generally, the velocity distribution analysis demonstrates that our simulation successfully reproduces both large- and small-scale features observed during this TSE. However, it also becomes evident that the simulation only partially reproduces the magnetic field configuration and low-speed characteristics on the left limb, likely due to missing real-time observational data on the left side.

For the thermal pressure distribution in Figure \ref{fig:05variable}b, it is noted that it has evolved into anisotropic distributions that correspond to the magnetic field configuration, which are different from the initial isothermal atmosphere condition which was only radially structured. There are several regions with relatively high thermal pressure corresponding to the regions with low radial velocity as PS1, PS2, PS3, PS4, S3, and S5 in Figure \ref{fig:05variable}a, and they also align with positions of pseudo-streamers and streamers in Figure \ref{fig:03maglines}d. Moreover, a low-pressure arc appears on the left side of the solar disk, approximately matching with the location of closed magnetic field lines at that position. Overall, these distributions indicate that the evolution of the thermal pressure is physically reasonable.

For the density distribution in Figure \ref{fig:05variable}d, it is seen that the density also exhibits certain distributions similar to the thermal pressure configuration in Figure \ref{fig:05variable}b, but overall is more isotropic. It suggests that, compared to the initial stratified atmosphere, no significant evolution in density has been found, although some small localized features according to the magnetic field have emerged. It is common that thermodynamic physical quantities in MHD simulations under the polytropic approximation, namely $\gamma=1.05$, would not undergo considerable variations during relaxation \citep{perri2022coconut1, kuzma2023coconut, mikic1999mhd}. However, it remains sensitive to other variables including magnetic field and velocities and can serve as a validation criterion of the simulations. 

We also calculate the plasma beta $\beta=p/(B^2/2\mu_0)$ on the meridian plane in Figure \ref{fig:05variable}c. It can be seen that $\beta$ remains reasonable in most regions, which is relatively large at the cusp location and above the streamers, and relatively small at regions with strong magnetic field strength. The contours of $\beta=1$ are found to be largely coincident with the outer boundary of the streamers and also extend toward their tops and above. Generally, the distribution of $\beta$ appears to be physical and accurately reflects the coronal conditions under the solar wind effects. However, due to the relatively weak magnetic field intensity of the outflow field model with $l_{max}=10$, the overall $\beta$ values tend to be slightly higher than expected.

\begin{figure*}
    \centering
    \includegraphics[width=\textwidth]{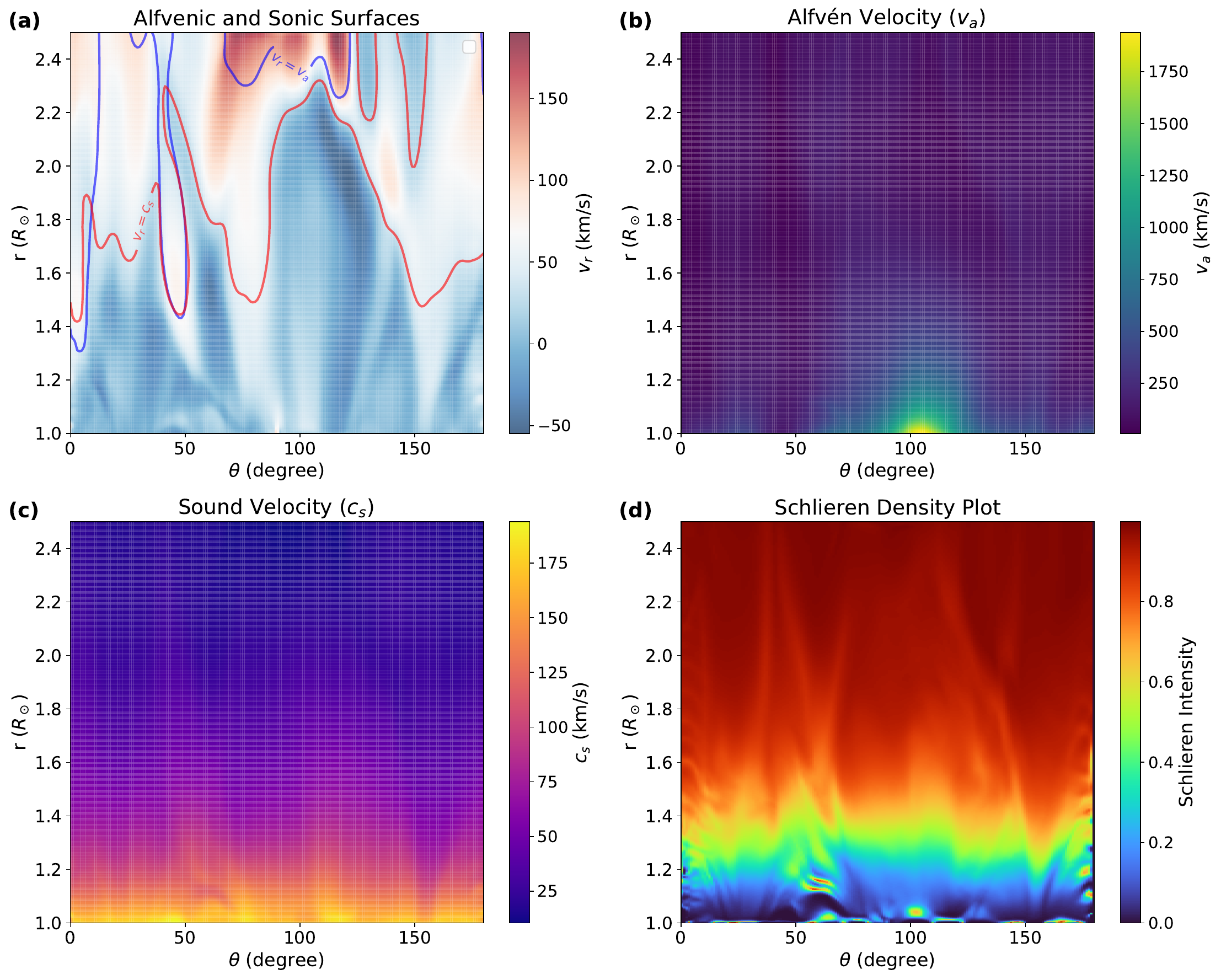}
    \caption{Distributions of (a) radial velocity $v_r$, (b) Alfvén velocity $v_a$, (c) sound velocity $c_s$, and (d) the schlieren density plot on a typical plane $\phi=130\degree$. The blue and red contours on (a) denote the position where $v_r=v_a$ and $v_r=c_s$.}
    \label{fig:supple:thermal}
\end{figure*}

Moreover, we analyze the derived physical features of coronal plasma. We calculate the Alfvén velocity, sound velocity, and make a direct comparison of them with the radial velocity in Figure \ref{fig:supple:thermal} on a selected typical plane $\phi=130\degree$. It is found that the overall Alfvén velocity is relatively small compared to the typical values in the low solar corona because of the reduced magnetic field strength. However, the magnitudes of both the Alfvén velocity and the sound velocity fall within a reasonable range under the model assumptions in Figures \ref{fig:supple:thermal}b and \ref{fig:supple:thermal}c. The overall distribution of the radial velocity is consistent with the influence of the solar wind, exhibiting low-velocity bulks and high-velocity streams at higher altitudes in Figure \ref{fig:supple:thermal}a. Moreover, the positions of the contours where $v_r = v_a$ and $v_r = c_s$ in Figure \ref{fig:supple:thermal}a are also consistent with general expectations, supporting the reliability of the simulation results. We also calculate the schlieren density plot on the same plane in Figure \ref{fig:supple:thermal}d, which visualizes spatial gradient of the density. Regions with sharp changes appear bright, while uniform areas remain dark. It effectively maps out the location and structure of shock waves, discontinuities, and other compressible features in the magnetized plasma. 
It can be seen that besides the increasing density gradient resulting from the stratified atmospheric density, compressional features also occur in closed-field regions with strong magnetic fields in the low corona, such as beneath streamers and pseudo-streamers.

\subsection{Pseudo Radiation Synthesis and Radiation Synthesis}
\label{sec:results:radiation}
In the MHD relaxation where we impose a lowered polytropic index, thermodynamic physical quantities including $\rho, ~p,$ and other variables are not completely reliable compared to MHD simulations with complicated heating mechanisms. Another important factor is that the bottom boundary magnetogram has been smoothed with $l_{max}=10$. Therefore, to further diagnose the simulation results, we pay attention to magnetic field relevant features, including the distributions of the current density, which allows for a more intuitive quantification of the results of this simulation. 

\begin{figure*}
    \centering
    \includegraphics[width=\textwidth]{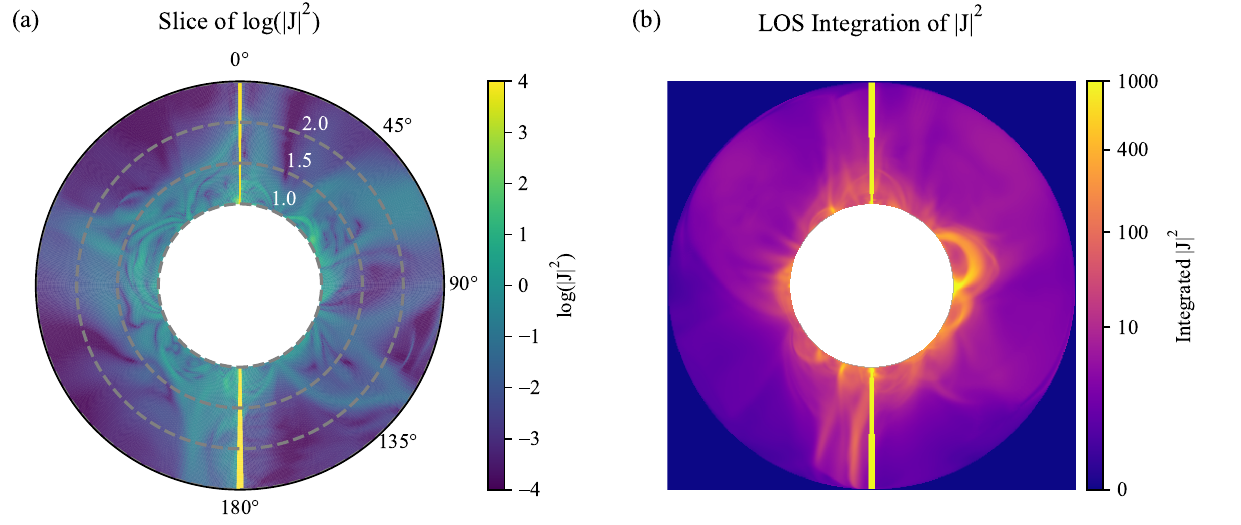}
    \caption{The meridian plane displays (a) the simulated current density distribution on this slice; (b) the simulated line-of-sight integration of the current intensity. The yellow artifact values on the lines near the polar regions are numerical caused by the processing of poles.}
    \label{fig:06current}
\end{figure*}

The simulated current density on the meridian plane when the TSE happened is in Figure \ref{fig:06current}a, while the line of sight integration of the current density is in Figure \ref{fig:06current}b. From Figure \ref{fig:06current}a, the current distribution is also found to exhibit distinct regional characteristics, mainly aligning with the positions of large streamers and pseudo-streamers. It forms several cusp-shaped structures, located on the east and the west limbs of the solar disk and polar regions. Within the cusps, the currents display complex patterns of closed magnetic fields, while extending outward to form large-scale current sheet structures. Outside the cusp regions, the current density remains generally weak. 

From Figure \ref{fig:06current}b, after integrating the current density along the line-of-sight, it is noted that more extensive and radiation-like structures appear, and this method is referred as the pseudo radiation synthesis. This is a helpful method to diagnose pseudo-radiative features of zero-$\beta$ \citep{li2025rbsl} and polytropic MHD simulations, and therefore is suitable for our MHD simulations with the energy equation and $\gamma=1.05$. It is seen that there is one evident closed loop in the southeast (lower left) limb, corresponding to the magnetic field loops L7 and L8 in Figure \ref{fig:03maglines}a, and the streamer S5 in the velocity map in Figure \ref{fig:05variable}a. There are also some other pseudo-streamer-like structures in the polar regions with similar positions of pseudo streamers in Figure \ref{fig:06current}a, and the overall structure exhibits more extensive and larger-scale characteristics. Moreover, it is noted that the extensive pseudo-streamer in the north pole is oriented westward, in contrast to the eastward orientation exhibited by the one in Figure \ref{fig:06current}a, which may be attributed to the influence of foreground structures on the current distribution during line-of-sight integration.

\begin{figure*}
    \centering
    \includegraphics[width=\textwidth]{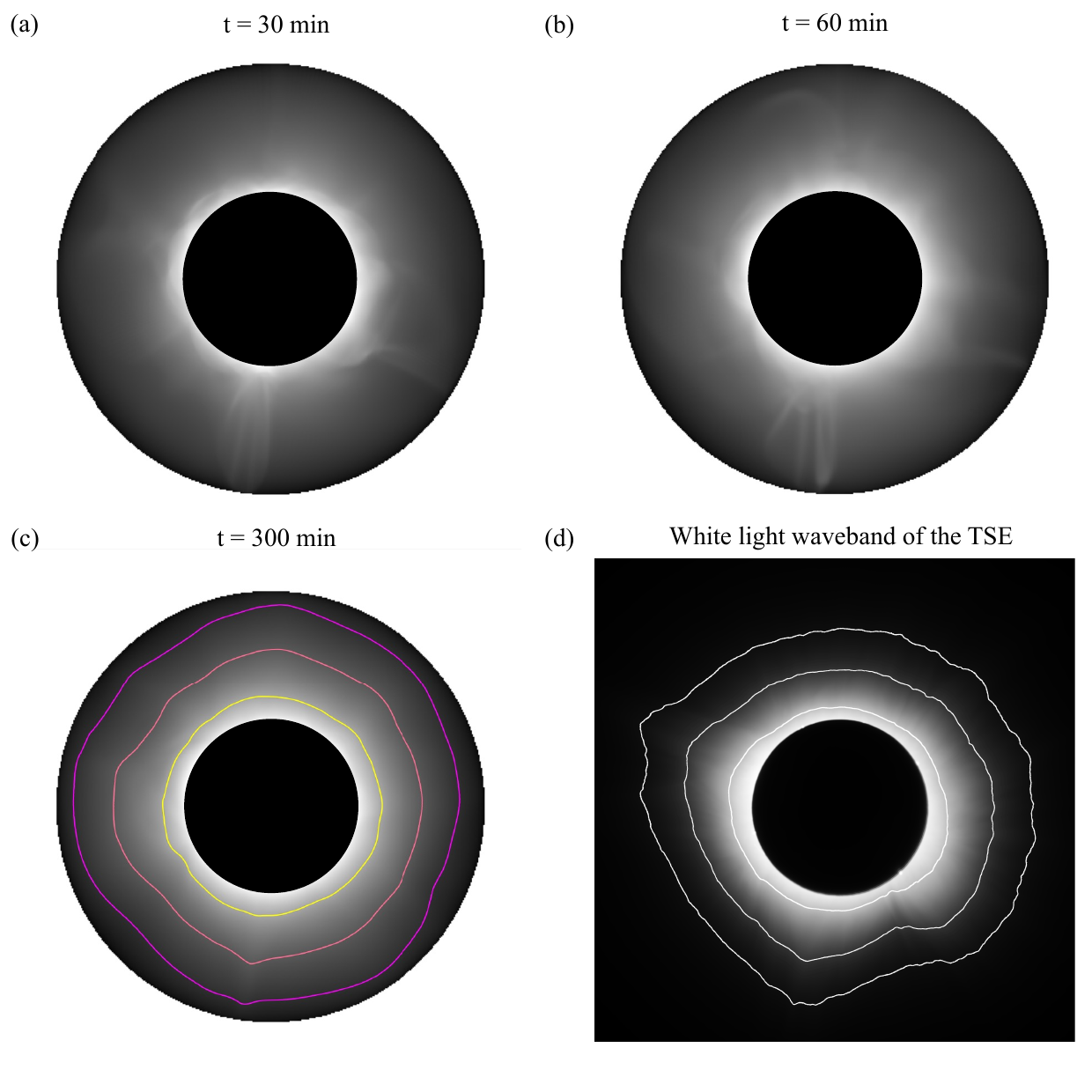}
    \caption{The meridian planes displaying the synthesized and observed white-light radiation. (a) The synthesized white-light radiation at the physical relaxation time $t=30$ minutes. (b) The radiation at the physical relaxation time $t=60$ minutes. (c) The radiation at the physical relaxation time $t=300$ minutes, which already reaches the steady state. The three contour levels are defined at 40\%, 50\%, 60\% of the log-scaled intensity values, respectively. (d) The original observed TSE image in the white-light waveband. The three lightness contours are brightness contours at $r=1.25, ~1.75,~2.25R_\odot$, respectively.}
    \label{fig:08radiation}
\end{figure*}

Furthermore, we synthesize the white-light radiation based on the MHD simulation results. The code we use to calculate synthesis radiation is the Radiation Synthesis Tools (RST)\citep{magwind2025}, and it is a visualization tool for efficiently analyzing MPI-AMRVAC simulation data and synthesize multi-wavelength radiation including extreme ultraviolet (EUV), white light, and H$\alpha$. RST can be used for both Cartesian coordinates and the spherical coordinates. The synthesis radiation and its comparison with the narrow-band TSE images are shown in Figure \ref{fig:08radiation}. In Figures \ref{fig:08radiation}a and \ref{fig:08radiation}b, the radiation during the relaxation process is displayed, at the physical time $t=30,~60$ minutes. It can be observed that during the early phase of the relaxation, the physical quantities including the density and the thermal pressure, which significantly affect the white-light emission, have not yet reached equilibrium and exhibit great variations, and therefore the synthesized radiation has more pronounced fluctuations. It is obvious that there already has been one pseudo streamer in the south pole, while another relatively large streamer appears in the southwestern direction. 

At the physical relaxation time $t=300$ minutes, our MHD evolution has already reached a steady state, and the density and temperature no longer undergo significant evolutions in Figure \ref{fig:08radiation}c. It reveals that the radiation at this stage is globally diffuse and extensive in the large scale due to the assumption of $\gamma=1.05$. From the contours of the luminosity, we can see that there are still multiple localized anisotropies in the radiation, such as the pseudo streamer located in the south pole and the closed coronal loop with comparatively weak radiation on the east limb. From the three contour levels at 40\%, 50\%, 60\% of the log-scaled intensity values, the locations of non-uniform radiation distributions can be distinctly identified. On the northeastern side of the solar disk, there are two strong radiation sources corresponding to PS4 in Figure \ref{fig:03maglines}d, and one is located near the south polar region to S5 and PS3 in Figure \ref{fig:03maglines}d. On the west limb of the solar disk, the bright structures corresponding to PS2 and S3 in Figure \ref{fig:03maglines}d can also be observed. The original white-light TSE observation displayed Figure \ref{fig:08radiation}d has more fine structures including the large streamers and plasma blobs. However, with the comparison of isophotes in Figures \ref{fig:08radiation}c and \ref{fig:08radiation}d, it is clear that the simulated synthetic radiation exhibits similarities with the observations, such as the stronger radiative flux in the southeastern locations and other characteristics on the western side, with discrepancies in the northeastern region. This suggests that the MHD simulation still has certain limitations, ultimately leading to suboptimal synthesis white-light radiation results. However, the synthetic white-light radiation still reveals certain large-scale features, and this method is a valuable tool for validating simulation results.

\section{Discussions and Summary} \label{sec:discussions}
We conduct a data-constrained MHD simulation of the global solar corona of the 2024 April 8 TSE with AMRVAC, and compare the simulation results with high-resolution observations of this TSE. Our simulation reproduces the magnetic field configuration and other physical variables with a fair amount of detail that resembles the observational views. We perform a comprehensive analysis of this simulation including the inspection of the magnetic field morphology and topology by calculating the QSL distribution, the comparison of the distributions of the coronal variables, and the calculation of the synthesized radiation and current density. The advantages and limitations of our simulations are discussed as follows.

Our model setup and numerical approaches have several key advantages which enhance and validate the reliability of the simulation results. First, for the initial magnetic field condition, we calculate the outflow field \citep{rice2021global} with modified Parker solar wind solution and boundary conditions. This model describes magnetic field configuration which is already relaxed with the solar wind effects under magneto-frictional method, which establishes a more accurate global magnetic field with a larger open flux, and thereby improves the efficiency of the subsequent MHD relaxation. Furthermore, we also compare different magnetograms and select the synchronic frames from SDO/HMI. These frames have relatively high spatial and temporal resolution, and the reconstructed magnetic field of the outflow field has a better consistency with the observations. 

Second, in this study, we solve the MHD equations with the energy equation under the AMRVAC framework. We have explored a comprehensive parameter space including different radial boundary conditions, divergence-fixing methods, and limiters to make sure the simulation results are reliable, physical, and relatively efficient. The numerical setup we currently adopt has been tested through multiple simulations, which are also consistent with previous studies \citep{wyper2021model, kuzma2023coconut, Cai2025MHDmodeling} and simultaneously yield the physical validity. Moreover, we use the GLM method to control the divergence of the magnetic field to ensure the numerical accuracy of the simulations. 

Third, we have systematically analyzed the simulation results and conducted detailed comparisons with high-quality TSE observations in the white-light wavelength and \ce{Fe XIV} emission taken by us. We examine the magnetic field morphology and topology with direct comparisons with TSE images and the calculation of QSLs. The results demonstrate that our simulations reproduce a majority of the magnetic field features including closed loops, streamers, pseudo streamers, and the open field. Moreover, the global coronal variables including the density, velocity and thermal pressure are reasonable and exhibit plasma variations and high-speed flows. Also, we conduct the pseudo radiation synthesis with the current density and white-light radiation synthesis with RST. The results demonstrate that the integrated current density can partially reproduce the radiation features with reasonable distributions, though this method remains a compromised approach although it is influenced by the foreground magnetic field. Additionally, although the synthesized white-light radiation is diffuse and lacking fine details, it still obtains large-scale structures of this TSE from the brightness contours, which validates the reliability of this MHD simulations.

However, this work has certain limitations, which we identify below and plan to explore in future researches. First, due to the observed limitations of magnetograms, the initial magnetic field configuration, which is the outflow field, did not completely reproduce an approximately accurate magnetic field distribution, especially in the far side. Furthermore, during the MHD relaxation, it is found that the final steady state is highly dependent on the initial magnetic field configuration, such as the positions and morphology of the streamers, pseudo streamers, and the open field. Therefore, the results of MHD simulations are highly sensitive to the input magnetograms at the bottom boundary and the global magnetic field model used. It emphasizes the importance of real-time, accurate, and multi-perspective observations of the solar photospheric magnetic field. Moreover, the radial boundary conditions we impose, especially at the top boundary, introduces certain limitations on the overall evolution of the magnetic field configuration. It sets the magnetic field direction to be radial on the source surface, therefore caps the upper limit of the streamers, and constrains global magnetic configurations to further expand. We intend to improve or fix this issue in subsequent studies by combining SCS models, and relaxing the global corona to $\sim 10-20~R_\odot$, while making full use of the AMR grid to simultaneously preserve both large- and small-scale features of the coupled flow and magnetic fields. However, such numerical setup stabilizes the simulation, maintains relatively high resolution of the inner corona and facilitates coupling with higher corona models.

The advantage of realizing this first global coronal model within the MPI-AMRVAC framework now allows us to further investigate the effects of different numerical algorithmic approaches, including the use of less diffusive limiters, the use of different divergence control measures, or the influence of employing multi-stage time-steppers of varying temporal order. Early experiments indicate that certain numerical schemes suffer from robustness issues and may lead to scenarios with plasma bulks with too high velocity near the HCS on the outer radial surface. We also note that AMRVAC primarily employs explicit methods, the relaxation process is relatively time-consuming. Achieving a steady-state solution for a given event typically requires considerable computational resources compared to codes utilizing implicit schemes. However, we can significantly save computational resources by selecting appropriate initial conditions such as the outflow field. Nevertheless, this approach simultaneously enables higher temporal resolution for future real-time solar wind relaxation simulations using AMRVAC.

It should also be noted that as this MHD simulation considers a simplified physical scenario, which solves the internal energy equation with a reduced polytropic index $\gamma=1.05$ without complicated heating mechanisms, the simulated thermodynamic quantities, including density and thermal pressure, inherently can only partially reproduce the observed radiative characteristics, as discussed in Section \ref{sec:results:radiation}. To better simulate the radiation in multi-wavelengths in the future, we need to consider more realistic heating mechanisms. In our current model, the boundary treatment allows both plasma and magnetic fields to interact across the boundary under sub-Alfvénic and subsonic surfaces, maintaining a physically consistent and numerically stable coupling with the solar wind. Nevertheless, a more rigorous implementation based on characteristic boundary conditions could further improve the subsonic-Alfvénic interfaces in global corona models by explicitly separating incoming and outgoing MHD wave components \citep{WU1987267,hayashi2005ApJSboundary, Hayashi2022ApJEinversion, Tarr2024ApJSdrivenBC}. A characteristic boundary setting is also more stable for data-driven boundaries. However, they are under developments and testing in AMRVAC, and therefore cannot yet be efficiently applied here. We aim to further develop such boundary treatments to achieve a more dynamic and accurate representation of the low-coronal evolution.

Considering the above factors, this study presents a successful implementation of a data-constrained global near-Sun coronal magnetic field relaxation model using AMRVAC, with solar wind effects constrained by observed magnetograms. First, we have simulated the global coronal configuration during this TSE event, capturing critical physical parameters including magnetic field topology, plasma velocity, and density distributions. Furthermore, we have quantitatively evaluated how different magnetogram inputs affect the simulation outcomes, and highlight the importance of real-time, multi-perspective magnetic field measurements in accurate coronal simulations. This generally establishes a reliable coronal model within $2.5~R_\odot$, which can be compared with observations of TSE and coronagraphs to examine the large-scale structure of the corona and the overall magnetic field. It also enables us to diagnose the velocity and density of the coronal plasma under solar wind effects. Second, this model establishes a physically accurate and consistent background solar corona profile, which is critical for simulations of the filaments and CMEs dynamics at active-region spatial scales. In particular, it can provide more realistic velocity and magnetic field backgrounds for the eruption of filaments or MFRs, especially suitable for studying how an eruptive MFR propagates to greater heights ($\sim2-3~R_\odot$) with quantified kinematics including the velocity and energy \citep{fan2024data}, and the morphological evolution \citep{Cai2025MHDmodeling}. However, the radiative characteristics are over simplified and need to be reproduced with more physical heating mechanisms in the future. Furthermore, it also enables to study magnetic reconnections between erupting structures and the ambient fields including open field and streamers under solar wind conditions, thus enabling to trace the origin regions of the fast and slow solar wind. 

In future studies, we will incorporate MFRs with flux rope embedding methods including the regularized Biot–Savart laws \citep[RBSL;][]{titov2018rbsl,guo2019rbsl,li2025rbsl, guo2024rbsl} and non-linear force free field extrapolation methods \citep{guo2017review, mackay2012models}, and facilitate the AMR capabilities of AMRVAC. This implementation will achieve a more computationally efficient numerical mesh, thereby optimizing computing resources. Moreover, due to the flexible architecture of multiple modules under AMRVAC, our coronal model can interface with various heliosphere and interplanetary models including semi-empirical and MHD models for space weather simulations and forecasting. We plan to extend the coronal model to $\sim 10-20~R_\odot$ with an SCS model, Wang-Sheeley-Arge (WSA) model and MHD relaxation, which have undergone several validation tests and are available in MagWind. Interplanetary models including EUHFORIA and ICARUS can be coupled with this coronal model to conduct a comprehensive investigation of magnetic structures and their coupled evolution with the solar wind from the solar surface out to several AUs. This can be used to investigate the properties of the solar wind and CMEs in the Sun-Earth space through comparisons with in-situ observations, extending the spatial domain of simulations and providing practical value for studying and predicting space weather.

\acknowledgments

We thank for the useful discussions with Y.W.N. and C.Z.. The observational data are provided courtesy of NASA/SDO and the AIA and HMI science teams. Y.H.L., Y.G., J.H.G., and M.D.D. are supported by the National Key R\&D Program of China (2022YFF0503004, 2021YFA1600504, and 2020YFC2201200) and NSFC (12333009). R.K. acknowledges funding from Internal KU Leuven funds in project UnderRadioSun (C16/24/010) and from FWO project Helioskill (G0B9923N). The numerical simulations are performed in the High Performance Computing Center (HPCC) at Nanjing University.


\bibliographystyle{apj}
\bibliography{solar_wind}

\begin{thebibliography}{98}
\expandafter\ifx\csname natexlab\endcsname\relax\def\natexlab#1{#1}\fi

\bibitem[{{Altschuler} \& {Newkirk}(1969)}]{Altschuler&Newkirk1969pfss}
{Altschuler}, M.~D. \& {Newkirk}, Jr., G. 1969, \solphys, 9, 131

\bibitem[{{Antiochos} {et~al.}(2011){Antiochos}, {Miki{\'c}}, {Titov},
  {Lionello}, \& {Linker}}]{antiochos2011solarwind}
{Antiochos}, S.~K., {Miki{\'c}}, Z., {Titov}, V.~S., {Lionello}, R., \&
  {Linker}, J.~A. 2011, \apj, 731, 112

\bibitem[{{Arge} {et~al.}(2010){Arge}, {Henney}, {Koller}, {Compeau}, {Young},
  {MacKenzie}, {Fay}, \& {Harvey}}]{2010AIPCAPADT}
{Arge}, C.~N., {Henney}, C.~J., {Koller}, J., {Compeau}, C.~R., {Young}, S.,
  {MacKenzie}, D., {Fay}, A., \& {Harvey}, J.~W. 2010, in American Institute of
  Physics Conference Series, Vol. 1216, Twelfth International Solar Wind
  Conference, ed. M.~{Maksimovic}, K.~{Issautier}, N.~{Meyer-Vernet},
  M.~{Moncuquet}, \& F.~{Pantellini} (AIP), 343--346

\bibitem[{{Aslanyan} {et~al.}(2024){Aslanyan}, {Scott}, {Wilkins}, {Meyer},
  {Pontin}, \& {Yeates}}]{Aslanyan2024fieldline}
{Aslanyan}, V., {Scott}, R.~B., {Wilkins}, C.~P., {Meyer}, K.~A., {Pontin},
  D.~I., \& {Yeates}, A.~R. 2024, \apj, 971, 137

\bibitem[{{Asvestari} {et~al.}(2024){Asvestari}, {Temmer}, {Caplan}, {Linker},
  {Heinemann}, {Pinto}, {Henney}, {Arge}, {Owens}, {Madjarska}, {Pomoell},
  {Hofmeister}, {Scolini}, \& {Samara}}]{asvestari2024openfield}
{Asvestari}, E., {Temmer}, M., {Caplan}, R.~M., {Linker}, J.~A., {Heinemann},
  S.~G., {Pinto}, R.~F., {Henney}, C.~J., {Arge}, C.~N., {Owens}, M.~J.,
  {Madjarska}, M.~S., {Pomoell}, J., {Hofmeister}, S.~J., {Scolini}, C., \&
  {Samara}, E. 2024, \apj, 971, 45

\bibitem[{{Badman} {et~al.}(2022){Badman}, {Brooks}, {Poirier}, {Warren},
  {Petrie}, {Rouillard}, {Nick Arge}, {Bale}, {de Pablos Ag{\"u}ero}, {Harra},
  {Jones}, {Kouloumvakos}, {Riley}, {Panasenco}, {Velli}, \&
  {Wallace}}]{badman2022constrain}
{Badman}, S.~T., {Brooks}, D.~H., {Poirier}, N., {Warren}, H.~P., {Petrie}, G.,
  {Rouillard}, A.~P., {Nick Arge}, C., {Bale}, S.~D., {de Pablos Ag{\"u}ero},
  D., {Harra}, L., {Jones}, S.~I., {Kouloumvakos}, A., {Riley}, P.,
  {Panasenco}, O., {Velli}, M., \& {Wallace}, S. 2022, \apj, 932, 135

\bibitem[{Baratashvili {et~al.}(2024)Baratashvili, Grison, Schmieder, Demoulin,
  \& Poedts}]{baratashvili2024multi}
Baratashvili, T., Grison, B., Schmieder, B., Demoulin, P., \& Poedts, S. 2024,
  arXiv preprint arXiv:2405.17988

\bibitem[{Baratashvili {et~al.}(2022)Baratashvili, Verbeke, Wijsen, \&
  Poedts}]{baratashvili2022improving}
Baratashvili, T., Verbeke, C., Wijsen, N., \& Poedts, S. 2022, Astronomy \&
  Astrophysics, 667, A133

\bibitem[{{Benavitz} {et~al.}(2024){Benavitz}, {Boe}, \&
  {Habbal}}]{Boe2024PFSS}
{Benavitz}, L.~F., {Boe}, B., \& {Habbal}, S.~R. 2024, \apj, 974, 178

\bibitem[{Bhowmik {et~al.}(2022)Bhowmik, Yeates, \&
  Rice}]{bhowmik2022exploring}
Bhowmik, P., Yeates, A., \& Rice, O. 2022, Solar Physics, 297, 41

\bibitem[{{Boe} {et~al.}(2020){Boe}, {Habbal}, \&
  {Druckm{\"u}ller}}]{boe2020TSE}
{Boe}, B., {Habbal}, S., \& {Druckm{\"u}ller}, M. 2020, \apj, 895, 123

\bibitem[{{Brchnelova} {et~al.}(2023){Brchnelova}, {Ku{\'z}ma}, {Zhang},
  {Lani}, \& {Poedts}}]{brchnelova2023A&Abeta}
{Brchnelova}, M., {Ku{\'z}ma}, B., {Zhang}, F., {Lani}, A., \& {Poedts}, S.
  2023, \aap, 676, A83

\bibitem[{{Cai} {et~al.}(2025){Cai}, {Zhang}, {Jiang}, {Yan}, {Feng}, {Zuo}, \&
  {Wang}}]{Cai2025MHDmodeling}
{Cai}, J., {Zhang}, L., {Jiang}, C., {Yan}, K., {Feng}, X., {Zuo}, P., \&
  {Wang}, Y. 2025, \mnras, 538, 2569

\bibitem[{{Chen}(2011)}]{chen2011review}
{Chen}, P.~F. 2011, Living Reviews in Solar Physics, 8, 1

\bibitem[{{Cheng} {et~al.}(2017){Cheng}, {Guo}, \& {Ding}}]{cheng2017review}
{Cheng}, X., {Guo}, Y., \& {Ding}, M. 2017, Science China Earth Sciences, 60,
  1383

\bibitem[{{Cranmer}(2004)}]{Cranmer2004lambertPKSW}
{Cranmer}, S.~R. 2004, American Journal of Physics, 72, 1397

\bibitem[{{Dedner} {et~al.}(2002){Dedner}, {Kemm}, {Kr{\"o}ner}, {Munz},
  {Schnitzer}, \& {Wesenberg}}]{dedner2002glm}
{Dedner}, A., {Kemm}, F., {Kr{\"o}ner}, D., {Munz}, C.~D., {Schnitzer}, T., \&
  {Wesenberg}, M. 2002, Journal of Computational Physics, 175, 645

\bibitem[{Downs {et~al.}(2025)Downs, Linker, Caplan, Mason, Riley, Davidson,
  Reyes, Palmerio, Lionello, Turtle, Ben-Nun, Stulajter, Titov, Török, Upton,
  Attie, Jha, Arge, Henney, Valori, Strecker, Calchetti, Germerott, Hirzberger,
  Suárez, Rodríguez, Solanki, Cheng, \& Wu}]{cooper2025science}
Downs, C., Linker, J.~A., Caplan, R.~M., Mason, E.~I., Riley, P., Davidson, R.,
  Reyes, A., Palmerio, E., Lionello, R., Turtle, J., Ben-Nun, M., Stulajter,
  M.~M., Titov, V.~S., Török, T., Upton, L.~A., Attie, R., Jha, B.~K., Arge,
  C.~N., Henney, C.~J., Valori, G., Strecker, H., Calchetti, D., Germerott, D.,
  Hirzberger, J., Suárez, D.~O., Rodríguez, J.~B., Solanki, S.~K., Cheng, X.,
  \& Wu, S. 2025, Science, 0, eadq0872

\bibitem[{{Druckm{\"u}ller}(2009)}]{2009ApJ...706.1605D}
{Druckm{\"u}ller}, M. 2009, \apj, 706, 1605

\bibitem[{{Druckm{\"u}ller} {et~al.}(2006){Druckm{\"u}ller}, {Ru{\v{s}}in}, \&
  {Minarovjech}}]{2006CoSka..36..131D}
{Druckm{\"u}ller}, M., {Ru{\v{s}}in}, V., \& {Minarovjech}, M. 2006,
  Contributions of the Astronomical Observatory Skalnate Pleso, 36, 131

\bibitem[{Fan(2016)}]{fan2016modeling}
Fan, Y. 2016, The Astrophysical Journal, 824, 93

\bibitem[{Fan(2017)}]{fan2017mhd}
---. 2017, The Astrophysical Journal, 844, 26

\bibitem[{{Fan}(2018)}]{fan2018mhd}
{Fan}, Y. 2018, \apj, 862, 54

\bibitem[{{Fan}(2022)}]{fan2022improved}
---. 2022, \apj, 941, 61

\bibitem[{{Fan} {et~al.}(2024){Fan}, {Kazachenko}, {Afanasyev}, \&
  {Fisher}}]{fan2024data}
{Fan}, Y., {Kazachenko}, M.~D., {Afanasyev}, A.~N., \& {Fisher}, G.~H. 2024,
  \apj, 975, 206

\bibitem[{{Feng} {et~al.}(2023){Feng}, {Lv}, {Xiang}, \&
  {Jiang}}]{Feng2023MNRAS.519.6297F}
{Feng}, X., {Lv}, J., {Xiang}, C., \& {Jiang}, C. 2023, \mnras, 519, 6297

\bibitem[{{Feng} {et~al.}(2015){Feng}, {Ma}, \& {Xiang}}]{Feng2015Data}
{Feng}, X., {Ma}, X., \& {Xiang}, C. 2015, Journal of Geophysical Research
  (Space Physics), 120, 10,159

\bibitem[{Gottlieb \& Shu(1998)}]{Gottlieb1998TotalVD}
Gottlieb, S. \& Shu, C.-W. 1998, Math. Comput., 67, 73

\bibitem[{{Groth} {et~al.}(2000){Groth}, {De Zeeuw}, {Gombosi}, \&
  {Powell}}]{Groth2000JGR...10525053G}
{Groth}, C. P.~T., {De Zeeuw}, D.~L., {Gombosi}, T.~I., \& {Powell}, K.~G.
  2000, \jgr, 105, 25053

\bibitem[{Guo {et~al.}(2024)Guo, Linan, Poedts, Guo, Lani, Schmieder,
  Brchnelova, Perri, Baratashvili, Ni, {et~al.}}]{guo2024modeling}
Guo, J., Linan, L., Poedts, S., Guo, Y., Lani, A., Schmieder, B., Brchnelova,
  M., Perri, B., Baratashvili, T., Ni, Y., {et~al.} 2024, Astronomy \&
  Astrophysics, 683, A54

\bibitem[{{Guo} {et~al.}(2024{\natexlab{a}}){Guo}, {Linan}, {Poedts}, {Guo},
  {Lani}, {Schmieder}, {Brchnelova}, {Perri}, {Baratashvili}, {Ni}, \&
  {Chen}}]{guo2024rbsl}
{Guo}, J.~H., {Linan}, L., {Poedts}, S., {Guo}, Y., {Lani}, A., {Schmieder},
  B., {Brchnelova}, M., {Perri}, B., {Baratashvili}, T., {Ni}, Y.~W., \&
  {Chen}, P.~F. 2024{\natexlab{a}}, \aap, 683, A54

\bibitem[{{Guo} {et~al.}(2023){Guo}, {Ni}, {Zhong}, {Guo}, {Xia}, {Li},
  {Poedts}, {Schmieder}, \& {Chen}}]{guo2023MFRbirth-death}
{Guo}, J.~H., {Ni}, Y.~W., {Zhong}, Z., {Guo}, Y., {Xia}, C., {Li}, H.~T.,
  {Poedts}, S., {Schmieder}, B., \& {Chen}, P.~F. 2023, \apjs, 266, 3

\bibitem[{{Guo} {et~al.}(2017){Guo}, {Cheng}, \& {Ding}}]{guo2017review}
{Guo}, Y., {Cheng}, X., \& {Ding}, M. 2017, Science China Earth Sciences, 60,
  1408

\bibitem[{Guo {et~al.}(2012)Guo, Ding, Liu, Sun, DeRosa, \&
  Wiegelmann}]{guo2012modeling}
Guo, Y., Ding, M., Liu, Y., Sun, X., DeRosa, M., \& Wiegelmann, T. 2012, The
  Astrophysical Journal, 760, 47

\bibitem[{{Guo} {et~al.}(2024{\natexlab{b}}){Guo}, {Guo}, {Ni}, {Xia}, {Zhong},
  {Ding}, {Chen}, \& {Keppens}}]{guo2024review}
{Guo}, Y., {Guo}, J., {Ni}, Y., {Xia}, C., {Zhong}, Z., {Ding}, M., {Chen}, P.,
  \& {Keppens}, R. 2024{\natexlab{b}}, Reviews of Modern Plasma Physics, 8, 29

\bibitem[{{Guo} {et~al.}(2019){Guo}, {Xu}, {Ding}, {Chen}, {Xia}, \&
  {Keppens}}]{guo2019rbsl}
{Guo}, Y., {Xu}, Y., {Ding}, M.~D., {Chen}, P.~F., {Xia}, C., \& {Keppens}, R.
  2019, \apjl, 884, L1

\bibitem[{{Guo} {et~al.}(2021){Guo}, {Zhong}, {Ding}, {Chen}, {Xia}, \&
  {Keppens}}]{guo2021rbslsphere}
{Guo}, Y., {Zhong}, Z., {Ding}, M.~D., {Chen}, P.~F., {Xia}, C., \& {Keppens},
  R. 2021, \apj, 919, 39

\bibitem[{Habbal {et~al.}(2014)Habbal, Druckmüller, \&
  Morgan}]{2014..8466..17}
Habbal, S., Druckmüller, M., \& Morgan, H. 2014, in , 17--24

\bibitem[{{Habbal} {et~al.}(2011){Habbal}, {Cooper}, {Daw}, {Ding},
  {Druckmuller}, {Esser}, {Johnson}, \& {Morgan}}]{habbal2011arXivTSE}
{Habbal}, S.~R., {Cooper}, J., {Daw}, A., {Ding}, A., {Druckmuller}, M.,
  {Esser}, R., {Johnson}, J., \& {Morgan}, H. 2011, arXiv e-prints,
  arXiv:1108.2323

\bibitem[{{Habbal} {et~al.}(2021){Habbal}, {Druckm{\"u}ller}, {Alzate}, {Ding},
  {Johnson}, {Starha}, {Hoderova}, {Boe}, {Constantinou}, \&
  {Arndt}}]{habbal2021ApJTSE}
{Habbal}, S.~R., {Druckm{\"u}ller}, M., {Alzate}, N., {Ding}, A., {Johnson},
  J., {Starha}, P., {Hoderova}, J., {Boe}, B., {Constantinou}, S., \& {Arndt},
  M. 2021, \apjl, 911, L4

\bibitem[{{Harvey} {et~al.}(1996){Harvey}, {Hill}, {Hubbard}, {Kennedy},
  {Leibacher}, {Pintar}, {Gilman}, {Noyes}, {Title}, {Toomre}, {Ulrich},
  {Bhatnagar}, {Kennewell}, {Marquette}, {Patron}, {Saa}, \&
  {Yasukawa}}]{1996SciGONG}
{Harvey}, J.~W., {Hill}, F., {Hubbard}, R.~P., {Kennedy}, J.~R., {Leibacher},
  J.~W., {Pintar}, J.~A., {Gilman}, P.~A., {Noyes}, R.~W., {Title}, A.~M.,
  {Toomre}, J., {Ulrich}, R.~K., {Bhatnagar}, A., {Kennewell}, J.~A.,
  {Marquette}, W., {Patron}, J., {Saa}, O., \& {Yasukawa}, E. 1996, Science,
  272, 1284

\bibitem[{{Hayashi}(2005)}]{hayashi2005ApJSboundary}
{Hayashi}, K. 2005, \apjs, 161, 480

\bibitem[{{Hayashi} {et~al.}(2021){Hayashi}, {Abbett}, {Cheung}, \&
  {Fisher}}]{Hayashi2021ApJMHDmf}
{Hayashi}, K., {Abbett}, W.~P., {Cheung}, M. C.~M., \& {Fisher}, G.~H. 2021,
  \apjs, 254, 1

\bibitem[{{Hayashi} {et~al.}(2022{\natexlab{a}}){Hayashi}, {Wu}, \&
  {Liou}}]{Hayashi2022ApJEinversion}
{Hayashi}, K., {Wu}, C.-C., \& {Liou}, K. 2022{\natexlab{a}}, \apj, 930, 60

\bibitem[{{Hayashi} {et~al.}(2022{\natexlab{b}}){Hayashi}, {Wu}, \&
  {Liou}}]{Hayashi2022ApJSteadystate}
---. 2022{\natexlab{b}}, \apj, 940, 82

\bibitem[{{Hayashi} {et~al.}(2023){Hayashi}, {Wu}, \&
  {Liou}}]{Hayashi2023ApJSinner}
---. 2023, \apjs, 268, 39

\bibitem[{Huang {et~al.}(2024)Huang, Toth, Huang, Sachdeva, van~der Holst, \&
  Manchester}]{huang2024adjustingpfss}
Huang, Z., Toth, G., Huang, J., Sachdeva, N., van~der Holst, B., \& Manchester,
  W.~B. 2024, arXiv preprint arXiv:2403.01712

\bibitem[{{Keppens} \& {Goedbloed}(1999)}]{keppens1999polytropic}
{Keppens}, R. \& {Goedbloed}, J.~P. 1999, \aap, 343, 251

\bibitem[{{Keppens} {et~al.}(2023){Keppens}, {Popescu Braileanu}, {Zhou},
  {Ruan}, {Xia}, {Guo}, {Claes}, \& {Bacchini}}]{rony2023mpi}
{Keppens}, R., {Popescu Braileanu}, B., {Zhou}, Y., {Ruan}, W., {Xia}, C.,
  {Guo}, Y., {Claes}, N., \& {Bacchini}, F. 2023, \aap, 673, A66

\bibitem[{Keppens {et~al.}(2021)Keppens, Teunissen, Xia, \&
  Porth}]{keppens2021mpi}
Keppens, R., Teunissen, J., Xia, C., \& Porth, O. 2021, Computers \&
  Mathematics with Applications, 81, 316

\bibitem[{Kruse {et~al.}(2021)Kruse, Heidrich-Meisner, \&
  Wimmer-Schweingruber}]{kruse2021pfssevaluation}
Kruse, M., Heidrich-Meisner, V., \& Wimmer-Schweingruber, R. 2021, Astronomy \&
  Astrophysics, 645, A83

\bibitem[{Kruse {et~al.}(2020)Kruse, Heidrich-Meisner, Wimmer-Schweingruber, \&
  Hauptmann}]{kruse2020pfsselliptic}
Kruse, M., Heidrich-Meisner, V., Wimmer-Schweingruber, R., \& Hauptmann, M.
  2020, Astronomy \& Astrophysics, 638, A109

\bibitem[{Ku{\'z}ma {et~al.}(2023)Ku{\'z}ma, Brchnelova, Perri, Baratashvili,
  Zhang, Lani, \& Poedts}]{kuzma2023coconut}
Ku{\'z}ma, B., Brchnelova, M., Perri, B., Baratashvili, T., Zhang, F., Lani,
  A., \& Poedts, S. 2023, The Astrophysical Journal, 942, 31

\bibitem[{Li \& Chen(2025)}]{magwind2025}
Li, Y. \& Chen, G. 2025, MagWind

\bibitem[{Li {et~al.}(2025)Li, Guo, Guo, Ding, Xia, \& Keppens}]{li2025rbsl}
Li, Y., Guo, Y., Guo, J., Ding, M.~D., Xia, C., \& Keppens, R. 2025,
  Data-constrained Magnetohydrodynamic Simulation of a Filament Eruption in a
  Decaying Active Region 13079 on a Global Scale

\bibitem[{Linan {et~al.}(2024)Linan, Baratashvili, Lani, Schmieder, Brchnelova,
  Guo, \& Poedts}]{linan2024cme}
Linan, L., Baratashvili, T., Lani, A., Schmieder, B., Brchnelova, M., Guo, J.,
  \& Poedts, S. 2024, arXiv preprint arXiv:2411.19340

\bibitem[{Linan {et~al.}(2023)Linan, Regnault, Perri, Brchnelova, Kuzma, Lani,
  Poedts, \& Schmieder}]{linan2023self}
Linan, L., Regnault, F., Perri, B., Brchnelova, M., Kuzma, B., Lani, A.,
  Poedts, S., \& Schmieder, B. 2023, Astronomy \& Astrophysics, 675, A101

\bibitem[{Linker {et~al.}(2017)Linker, Caplan, Downs, Riley, Mikic, Lionello,
  Henney, Arge, Liu, Derosa, {et~al.}}]{linker2017open}
Linker, J., Caplan, R., Downs, C., Riley, P., Mikic, Z., Lionello, R., Henney,
  C., Arge, C., Liu, Y., Derosa, M., {et~al.} 2017, The Astrophysical Journal,
  848, 70

\bibitem[{{Linker} {et~al.}(2011){Linker}, {Lionello}, {Miki{\'c}}, {Titov}, \&
  {Antiochos}}]{linker2011openflux}
{Linker}, J.~A., {Lionello}, R., {Miki{\'c}}, Z., {Titov}, V.~S., \&
  {Antiochos}, S.~K. 2011, \apj, 731, 110

\bibitem[{{Lionello} {et~al.}(2023){Lionello}, {Downs}, {Mason}, {Linker},
  {Caplan}, {Riley}, {Titov}, \& {DeRosa}}]{Lionello2023MAS}
{Lionello}, R., {Downs}, C., {Mason}, E.~I., {Linker}, J.~A., {Caplan}, R.~M.,
  {Riley}, P., {Titov}, V.~S., \& {DeRosa}, M.~L. 2023, \apj, 959, 77

\bibitem[{{Liu} {et~al.}(2025){Liu}, {Liu}, {Manchester}, {Welling},
  {T{\'o}th}, {Gombosi}, {DeRosa}, {Bertello}, {Pevtsov}, {Pevtsov}, {Reardon},
  {Wilbanks}, {Rewoldt}, \& {Zhao}}]{Liu2025arXivforecast}
{Liu}, X., {Liu}, W., {Manchester}, IV, W.~B., {Welling}, D.~T., {T{\'o}th},
  G., {Gombosi}, T.~I., {DeRosa}, M.~L., {Bertello}, L., {Pevtsov}, A.~A.,
  {Pevtsov}, A.~A., {Reardon}, K., {Wilbanks}, K., {Rewoldt}, A., \& {Zhao}, L.
  2025, arXiv e-prints, arXiv:2503.10974

\bibitem[{{Mackay} \& {van Ballegooijen}(2005)}]{mackay2005model}
{Mackay}, D.~H. \& {van Ballegooijen}, A.~A. 2005, \apjl, 621, L77

\bibitem[{{Mackay} \& {Yeates}(2012{\natexlab{a}})}]{mackay2012models}
{Mackay}, D.~H. \& {Yeates}, A.~R. 2012{\natexlab{a}}, Living Reviews in Solar
  Physics, 9, 6

\bibitem[{{Mackay} \& {Yeates}(2012{\natexlab{b}})}]{mackay2012optimation}
---. 2012{\natexlab{b}}, Living Reviews in Solar Physics, 9, 6

\bibitem[{{Miki{\'c}} {et~al.}(2018){Miki{\'c}}, {Downs}, {Linker}, {Caplan},
  {Mackay}, {Upton}, {Riley}, {Lionello}, {T{\"o}r{\"o}k}, {Titov}, {Wijaya},
  {Druckm{\"u}ller}, {Pasachoff}, \& {Carlos}}]{mikic2018predict}
{Miki{\'c}}, Z., {Downs}, C., {Linker}, J.~A., {Caplan}, R.~M., {Mackay},
  D.~H., {Upton}, L.~A., {Riley}, P., {Lionello}, R., {T{\"o}r{\"o}k}, T.,
  {Titov}, V.~S., {Wijaya}, J., {Druckm{\"u}ller}, M., {Pasachoff}, J.~M., \&
  {Carlos}, W. 2018, Nature Astronomy, 2, 913

\bibitem[{{Miki{\'c}} \& {Linker}(1996)}]{mikic1996large-scale}
{Miki{\'c}}, Z. \& {Linker}, J.~A. 1996, in American Institute of Physics
  Conference Series, Vol. 382, Proceedings of the eigth International solar
  wind Conference: Solar wind eight, ed. D.~{Winterhalter}, J.~T. {Gosling},
  S.~R. {Habbal}, W.~S. {Kurth}, \& M.~{Neugebauer}, 104--107

\bibitem[{{Miki{\'c}} {et~al.}(1999){Miki{\'c}}, {Linker}, {Schnack},
  {Lionello}, \& {Tarditi}}]{mikic1999mhd}
{Miki{\'c}}, Z., {Linker}, J.~A., {Schnack}, D.~D., {Lionello}, R., \&
  {Tarditi}, A. 1999, Physics of Plasmas, 6, 2217

\bibitem[{{Narechania} {et~al.}(2021){Narechania}, {Nikoli{\'c}}, {Freret}, {De
  Sterck}, \& {Groth}}]{Narechania2021solarwind}
{Narechania}, N.~M., {Nikoli{\'c}}, L., {Freret}, L., {De Sterck}, H., \&
  {Groth}, C. P.~T. 2021, Journal of Space Weather and Space Climate, 11, 8

\bibitem[{Nikoli{\'c}(2019)}]{nikolic2019pfsssolutions}
Nikoli{\'c}, L. 2019, Space Weather, 17, 1293

\bibitem[{Perri {et~al.}(2023)Perri, Ku{\'z}ma, Brchnelova, Baratashvili,
  Zhang, Leitner, Lani, \& Poedts}]{perri2023coconut2}
Perri, B., Ku{\'z}ma, B., Brchnelova, M., Baratashvili, T., Zhang, F., Leitner,
  P., Lani, A., \& Poedts, S. 2023, The Astrophysical Journal, 943, 124

\bibitem[{Perri {et~al.}(2022)Perri, Leitner, Brchnelova, Baratashvili,
  Ku{\'z}ma, Zhang, Lani, \& Poedts}]{perri2022coconut1}
Perri, B., Leitner, P., Brchnelova, M., Baratashvili, T., Ku{\'z}ma, B., Zhang,
  F., Lani, A., \& Poedts, S. 2022, The Astrophysical Journal, 936, 19

\bibitem[{Pomoell \& Poedts(2018)}]{pomoell2018euhforia}
Pomoell, J. \& Poedts, S. 2018, Journal of Space Weather and Space Climate, 8,
  A35

\bibitem[{Porth {et~al.}(2014)Porth, Xia, Hendrix, Moschou, \&
  Keppens}]{porth2014mpi}
Porth, O., Xia, C., Hendrix, T., Moschou, S., \& Keppens, R. 2014, The
  Astrophysical Journal Supplement Series, 214, 4

\bibitem[{Rice \& Yeates(2021)}]{rice2021global}
Rice, O.~E. \& Yeates, A.~R. 2021, The Astrophysical Journal, 923, 57

\bibitem[{Riley {et~al.}(2012)Riley, Linker, Lionello, \& Mikic}]{RILEY20121}
Riley, P., Linker, J.~A., Lionello, R., \& Mikic, Z. 2012, Journal of
  Atmospheric and Solar-Terrestrial Physics, 83, 1, corotating Interaction
  Regions from Sun to Earth: Modeling their formation, evolution and
  geoeffectiveness

\bibitem[{{Schatten} {et~al.}(1969){Schatten}, {Wilcox}, \&
  {Ness}}]{Schatten1969pfss}
{Schatten}, K.~H., {Wilcox}, J.~M., \& {Ness}, N.~F. 1969, \solphys, 6, 442

\bibitem[{{Scherrer} {et~al.}(2012){Scherrer}, {Schou}, {Bush}, {Kosovichev},
  {Bogart}, {Hoeksema}, {Liu}, {Duvall}, {Zhao}, {Title}, {Schrijver},
  {Tarbell}, \& {Tomczyk}}]{2012SoPhHMI}
{Scherrer}, P.~H., {Schou}, J., {Bush}, R.~I., {Kosovichev}, A.~G., {Bogart},
  R.~S., {Hoeksema}, J.~T., {Liu}, Y., {Duvall}, T.~L., {Zhao}, J., {Title},
  A.~M., {Schrijver}, C.~J., {Tarbell}, T.~D., \& {Tomczyk}, S. 2012, \solphys,
  275, 207

\bibitem[{{Shi} {et~al.}(2024){Shi}, {Feng}, {Ying}, {Li}, \&
  {Gan}}]{Shi2024ApJglobal}
{Shi}, G., {Feng}, L., {Ying}, B., {Li}, S., \& {Gan}, W. 2024, \apj, 970, 131

\bibitem[{{Tarr} {et~al.}(2024){Tarr}, {Kee}, {Linton}, {Schuck}, \&
  {Leake}}]{Tarr2024ApJSdrivenBC}
{Tarr}, L.~A., {Kee}, N.~D., {Linton}, M.~G., {Schuck}, P.~W., \& {Leake},
  J.~E. 2024, \apjs, 270, 30

\bibitem[{{Titov} {et~al.}(2018){Titov}, {Downs}, {Miki{\'c}}, {T{\"o}r{\"o}k},
  {Linker}, \& {Caplan}}]{titov2018rbsl}
{Titov}, V.~S., {Downs}, C., {Miki{\'c}}, Z., {T{\"o}r{\"o}k}, T., {Linker},
  J.~A., \& {Caplan}, R.~M. 2018, \apjl, 852, L21

\bibitem[{{Titov} {et~al.}(2011){Titov}, {Miki{\'c}}, {Linker}, {Lionello}, \&
  {Antiochos}}]{Titov2011ApJtopology}
{Titov}, V.~S., {Miki{\'c}}, Z., {Linker}, J.~A., {Lionello}, R., \&
  {Antiochos}, S.~K. 2011, \apj, 731, 111

\bibitem[{{T{\"o}r{\"o}k} {et~al.}(2018){T{\"o}r{\"o}k}, {Downs}, {Linker},
  {Lionello}, {Titov}, {Miki{\'c}}, {Riley}, {Caplan}, \&
  {Wijaya}}]{torok2018sun2earth}
{T{\"o}r{\"o}k}, T., {Downs}, C., {Linker}, J.~A., {Lionello}, R., {Titov},
  V.~S., {Miki{\'c}}, Z., {Riley}, P., {Caplan}, R.~M., \& {Wijaya}, J. 2018,
  \apj, 856, 75

\bibitem[{T{\'o}th {et~al.}(2011)T{\'o}th, Van~der Holst, \&
  Huang}]{toth2011pfss}
T{\'o}th, G., Van~der Holst, B., \& Huang, Z. 2011, The Astrophysical Journal,
  732, 102

\bibitem[{{Usmanov} {et~al.}(2014){Usmanov}, {Goldstein}, \&
  {Matthaeus}}]{Usmanov2014threefluid}
{Usmanov}, A.~V., {Goldstein}, M.~L., \& {Matthaeus}, W.~H. 2014, \apj, 788, 43

\bibitem[{{Usmanov} {et~al.}(2018){Usmanov}, {Matthaeus}, {Goldstein}, \&
  {Chhiber}}]{Usmanov2018SteadySW}
{Usmanov}, A.~V., {Matthaeus}, W.~H., {Goldstein}, M.~L., \& {Chhiber}, R.
  2018, \apj, 865, 25

\bibitem[{{van der Holst} \& {Keppens}(2007)}]{vanrony2007hybridAMR}
{van der Holst}, B. \& {Keppens}, R. 2007, Journal of Computational Physics,
  226, 925

\bibitem[{{van der Holst} {et~al.}(2010){van der Holst}, {Manchester},
  {Frazin}, {V{\'a}squez}, {T{\'o}th}, \& {Gombosi}}]{van2010awsom}
{van der Holst}, B., {Manchester}, IV, W.~B., {Frazin}, R.~A., {V{\'a}squez},
  A.~M., {T{\'o}th}, G., \& {Gombosi}, T.~I. 2010, \apj, 725, 1373

\bibitem[{{van der Holst} {et~al.}(2014){van der Holst}, {Sokolov}, {Meng},
  {Jin}, {Manchester}, {T{\'o}th}, \& {Gombosi}}]{Van2014AWSoM}
{van der Holst}, B., {Sokolov}, I.~V., {Meng}, X., {Jin}, M., {Manchester}, IV,
  W.~B., {T{\'o}th}, G., \& {Gombosi}, T.~I. 2014, \apj, 782, 81

\bibitem[{Verbeke {et~al.}(2022)Verbeke, Baratashvili, \&
  Poedts}]{verbeke2022icarus}
Verbeke, C., Baratashvili, T., \& Poedts, S. 2022, Astronomy \& Astrophysics,
  662, A50

\bibitem[{Wagner {et~al.}(2022)Wagner, Asvestari, Temmer, Heinemann, \&
  Pomoell}]{wagner2022validation}
Wagner, A., Asvestari, E., Temmer, M., Heinemann, S., \& Pomoell, J. 2022,
  Astronomy \& Astrophysics, 657, A117

\bibitem[{{Wang} {et~al.}(2025){Wang}, {Yang}, {Poedts}, {Lani}, {Zhou}, {Gao},
  {Linan}, {Lv}, {Baratashvili}, {Guo}, {Lin}, {Su}, {Li}, {Zhang}, {Wei},
  {Yang}, {Li}, {Ma}, {Husidic}, {Jeong}, {Najafi-Ziyazi}, {Wang}, \&
  {Schmieder}}]{wang2025ApJSsip}
{Wang}, H., {Yang}, L., {Poedts}, S., {Lani}, A., {Zhou}, Y., {Gao}, Y.,
  {Linan}, L., {Lv}, J., {Baratashvili}, T., {Guo}, J., {Lin}, R., {Su}, Z.,
  {Li}, C., {Zhang}, M., {Wei}, W., {Yang}, Y., {Li}, Y., {Ma}, X., {Husidic},
  E., {Jeong}, H.-J., {Najafi-Ziyazi}, M., {Wang}, J., \& {Schmieder}, B. 2025,
  \apjs, 278, 59

\bibitem[{{Wiegelmann} {et~al.}(2017){Wiegelmann}, {Petrie}, \&
  {Riley}}]{wiegelmann2017coronalmagnetic}
{Wiegelmann}, T., {Petrie}, G. J.~D., \& {Riley}, P. 2017, \ssr, 210, 249

\bibitem[{Wu \& Wang(1987)}]{WU1987267}
Wu, S. \& Wang, J. 1987, Computer Methods in Applied Mechanics and Engineering,
  64, 267

\bibitem[{Wyper {et~al.}(2024)Wyper, Lynch, DeVore, Kumar, Antiochos, \&
  Daldorff}]{wyper2024model}
Wyper, P., Lynch, B., DeVore, C., Kumar, P., Antiochos, S., \& Daldorff, L.
  2024, The Astrophysical Journal, 975, 168

\bibitem[{Wyper {et~al.}(2021)Wyper, Antiochos, DeVore, Lynch, Karpen, \&
  Kumar}]{wyper2021model}
Wyper, P.~F., Antiochos, S.~K., DeVore, C.~R., Lynch, B.~J., Karpen, J.~T., \&
  Kumar, P. 2021, The Astrophysical Journal, 909, 54

\bibitem[{Xia {et~al.}(2018)Xia, Teunissen, El~Mellah, Chan{\'e}, \&
  Keppens}]{xia2018mpi}
Xia, C., Teunissen, J., El~Mellah, I., Chan{\'e}, E., \& Keppens, R. 2018, The
  Astrophysical Journal Supplement Series, 234, 30

\bibitem[{{Yeates} {et~al.}(2018){Yeates}, {Amari}, {Contopoulos}, {Feng},
  {Mackay}, {Miki{\'c}}, {Wiegelmann}, {Hutton}, {Lowder}, {Morgan}, {Petrie},
  {Rachmeler}, {Upton}, {Canou}, {Chopin}, {Downs}, {Druckm{\"u}ller},
  {Linker}, {Seaton}, \& {T{\"o}r{\"o}k}}]{yeates2018SSRv..214...99Y}
{Yeates}, A.~R., {Amari}, T., {Contopoulos}, I., {Feng}, X., {Mackay}, D.~H.,
  {Miki{\'c}}, Z., {Wiegelmann}, T., {Hutton}, J., {Lowder}, C.~A., {Morgan},
  H., {Petrie}, G., {Rachmeler}, L.~A., {Upton}, L.~A., {Canou}, A., {Chopin},
  P., {Downs}, C., {Druckm{\"u}ller}, M., {Linker}, J.~A., {Seaton}, D.~B., \&
  {T{\"o}r{\"o}k}, T. 2018, \ssr, 214, 99

\bibitem[{{Zhang} {et~al.}(2022){Zhang}, {Chen}, {Liu}, \&
  {Wang}}]{fastqsl2021}
{Zhang}, P., {Chen}, J., {Liu}, R., \& {Wang}, C. 2022, \apj, 937, 26

\end{thebibliography}

\end{document}